\documentclass[twocolumn]{aastex62}

\newcommand{\Lya}{Ly$\alpha$}
\newcommand{\OII}{[O\,{\sc ii}]}
\newcommand{\OIII}{[O\,{\sc iii}]}

\newcommand{\Hb}{H$\beta$}

\newcommand{\NeIII}{[Ne\,{\sc iii}]}
\newcommand{\fesc}{$f_{\rm esc}$}

\newcommand{\xiion}{$\xi_{\rm ion}$}
\newcommand{\xiionzero}{$\xi_{\rm ion,0}$}

\newcommand{\kms}{km\,s$^{-1}$}

\newcommand{\HI}{H{\sc i}}
\newcommand{\HII}{H{\sc ii}}

\newcommand{\ebv}{E(B$-$V)}

\newcommand{\Msun}{$M_{\odot}$}

\defcitealias{fletcher2019}{Paper~I}

\received{September 15, 2019}
\revised{--}
\accepted{December 28, 2019}
\submitjournal{ApJ}

\shorttitle{LACES II: \fesc\ vs. optical spectroscopic properties}
\shortauthors{Nakajima et al.}

\begin{document}

\title{The Lyman Continuum Escape Survey - II: Ionizing Radiation as a Function of the \OIII$/$\OII\ Line Ratio}

\correspondingauthor{Kimihiko Nakajima}
\email{knakajima@nbi.ku.dk}

\author[0000-0003-2965-5070]{Kimihiko Nakajima}
\affiliation{Niels Bohr Institute, University of Copenhagen,
Lyngbyvej 2, 2100 Copenhagen \O, Denmark}
\affiliation{Cosmic DAWN Center}

\author[0000-0001-7782-7071]{Richard S. Ellis}
\affiliation{Department of Physics and Astronomy, University College London, 
Gower Street, London WC1E 6BT, UK}

\author[0000-0002-4271-0364]{Brant E. Robertson}
\affiliation{Department of Astronomy \& Astrophysics, University of California, Santa Cruz,
1156 High St., Santa Cruz, CA 95064, USA}

\author{Mengtao Tang}
\affiliation{Steward Observatory, University of Arizona, 933 N. Cherry Ave., Tucson, AZ 85721, USA}

\author{Daniel P. Stark}
\affiliation{Steward Observatory, University of Arizona, 933 N. Cherry Ave., Tucson, AZ 85721, USA}



\begin{abstract}

We discuss the rest-frame optical emission line spectra of a large 
($\sim 50$) sample of $z\sim 3.1$ Lyman alpha emitters (LAEs) whose
physical properties suggest such sources are promising analogs of galaxies
in the reionization era. Reliable Lyman continuum escape fractions have now 
been determined for a large sample of such LAEs from the Lyman Continuum Escape Survey 
(LACES) undertaken via deep HST imaging in the SSA22 survey area reported in 
\citet{fletcher2019}.
Using new measures of \OII\ emission secured from Keck
MOSFIRE spectra we re-examine, for a larger sample, earlier claims that Lyman continuum
leakages may correlate with the nebular emission line ratio \OIII$/$\OII\ as expected for 
density-bound \HII\ regions. We find that a large \OIII$/$\OII\  line ratio is indeed a necessary
condition for Lyman continuum leakage, strengthening earlier claims made using smaller
samples at various redshifts. However, not all LAEs with large \OIII$/$\OII\ line ratios
are leakers and leaking radiation appears not to be associated with
differences in other spectral diagnostics. This suggests the detection of leaking radiation
is modulated by an additional property, possibly the viewing angle for porous \HII\ regions.
We discuss our new results in the context of the striking bimodality of LAE leakers and 
non-leakers found in the LACES program and the implications for the sources of cosmic reionization.
\end{abstract}



\section{Introduction} \label{sec:intro}

The physical conditions that permit the leakage of ionizing radiation from 
star-forming galaxies is a topic of great interest. Recent analyses of the 
demographics and stellar properties of galaxies in the reionization era 
beyond a redshift $z\simeq6$ suggest a fraction of $10-20$\%\ of Lyman continuum 
photons must escape a typical low mass galaxy if such sources govern the 
process of cosmic reionization
\citep{robertson2013,bouwens2015_reionization,stark2016,DF2018}.
Since direct measures of  Lyman continuum (LyC) leakage are 
not possible at high redshift due to foreground IGM absorption, most recent 
work has focused on measures of the LyC escape fraction in low redshift analogs 
(e.g. \citealt{vanzella2015,siana2015,shapley2016,marchi2017,rutkowski2017,naidu2018,steidel2018,fletcher2019}).

Lyman alpha emitting galaxies (LAEs) are thought to be the most promising low 
redshift analogs of sources in the reionization era on account of their low 
gas-phase metallicity and high star formation rate. Ground-based spectroscopy 
reveals that many have intense \OIII\ emission \citep{nakajima2016,trainor2016}, 
a property which is inferred indirectly from Spitzer photometry for sources at  $z>6$
\citep{smit2015,roberts-borsani2016}. The Lyman Continuum Escape 
Survey (LACES) was designed to image a sample of $61$ $z=3.1$ LAEs found 
using narrow-band Subaru imaging in the SSA22 field 
\citep{hayashino2004,matsuda2005,yamada2012}
using a broad-band F336W filter with the Wide Field Camera 3 (WFC3) onboard 
Hubble Space Telescope (HST; GO 14747, PI: Robertson). 
In our first paper in this series 
(\citealt{fletcher2019}, hereafter \citetalias{fletcher2019}), on the basis of 
spectral energy distribution (SED) fitting, we presented convincing evidence for 
large escape fractions (\fesc\ $\sim 15-60$\%) for individual LAEs for $20$\%\ 
of the sample, in contrast to strict upper limits for the remainder. 
We found no strong correlation between this diversity of LyC radiation and 
other source properties such as stellar mass, UV luminosity and the equivalent widths 
of \OIII\ and Lyman alpha. We speculated on the origin of this curious bimodality 
in the emergence of ionizing radiation.

The inter-dependence of \fesc\ and nebular line emission was discussed 
by \citet{NO2014} in terms of their photoionisation models 
(see also \citealt{JO2013}). They found 
a possible correlation using the emission line ratio \OIII$/$\OII\ (hereafter O32) 
which was interpreted in terms of `density-bound' \HII\ regions. 
In contrast with `ionization-bound' \HII\ regions where LyC photons are fully 
absorbed within the radius of the Stromgren sphere, unusually high values of 
O32 would reflect partially-incomplete \HII\ regions where some LyC photons 
could escape. In this respect, therefore, LAEs would be powerful sources capable 
of driving cosmic reionization
(see also \citealt{marchi2018}). 
At the time of submission of \citetalias{fletcher2019}, 
a high fraction of the $61$ LACES sources had coverage of \OIII\ emission from 
several Keck MOSFIRE campaigns \citep{nakajima2016} but the coverage of \OII\ 
was limited. Accordingly, we have secured new MOSFIRE data improving the coverage
of \OII\ emission across the LACES sample so we can test for the expected 
trend between O32 and \fesc\ predicted by \citet{NO2014}.

A plan of the paper follows. In \S\ref{sec:data} we discuss the new spectroscopic data, its 
reduction and estimates of \OII\ emission and hence the O32 ratio. 
In \S\ref{sec:analysis} we revisit the LACES correlations in the context of our new line ratios
as well as the strength of the ionizing radiation field. We discuss the results 
in the context of the bimodality of LyC leakage found in \citetalias{fletcher2019} in 
\S\ref{sec:discussion}. 
Throughout the paper we adopt a concordance cosmology with 
$\Omega_\Lambda$=0.7, $\Omega_M$=0.3 and $H_0$=70 kms sec$^{-1}$ 
Mpc$^{-1}$.

\section{Data} \label{sec:data}

\subsection{MOSFIRE Observations and Data reduction} \label{ssec:data_observations}

Early MOSFIRE observations undertaken in the LACES area were described in
\citet{nakajima2016} and \citetalias{fletcher2019}. As a pilot observation, \citet{nakajima2016} 
discuss data for one MOSFIRE pointing (referred to here as mask\,1),  spectroscopically
covered in the K-band (sampling \OIII\ and \Hb)  and the H-band (sampling \OII\ and \NeIII).
HST/F336W coverage of LACES was determined in part based on this pilot observation.
In \citetalias{fletcher2019}, additional K-band spectroscopy for three further MOSFIRE pointings 
was presented (masks\,2--4), one of which (mask\,2) was also sampled in the H-band.
These additional pointings were chosen to include as many LACES sources as possible
with minimal overlap with mask\,1. Mask\,4 covered almost the same area as mask\,2, 
and was designed primarily to increase the depth for those sources whose K-band spectra 
were of low signal/noise.

In this paper we present MOSFIRE data from a further pointing (mask\,5) undertaken via a 
long integration in the H-band with the specific goal of improving the coverage of \OII\ emission
for sources well-studied in the K-band (i.e. \OIII) in masks\,2--4. The new H-band observations were 
taken in two second-half nights on UT August 3 and 4  2018 in clear conditions with a 
seeing of 0.4--0.8 arcsec. Observations were conducted in a similar manner to those reported
earlier, adopting a slit width of 0.7 arcsec and individual exposure times of 120\,sec with an AB
nod sequence of 3 arcsec separation. The total integration time for mask\,5 was $4.6$ hr.
Table \ref{tbl:summary_observations} provides a summary of our near-infrared spectroscopic campaign
of the LACES sample.

Data reduction was performed using the MOSFIRE DRP%
\footnote{
\url{https://keck-datareductionpipelines.github.io/MosfireDRP}}
in the manner described in \citet{nakajima2016}.
All spectroscopic data listed in Table \ref{tbl:summary_observations} were re-reduced with the latest (2018) 
version of MOSFIRE DRP. Briefly, the processing includes flat fielding, wavelength calibration, background subtraction 
and combining the nod positions. Wavelength calibration in the H-band was performed using OH 
sky lines and in the K-band via a combination of OH lines and Neon arcs. 
Flux calibration and telluric absorption corrections were obtained from A0V 
Hipparcos stars observed on the same night under the same seeing conditions, 
at similar air masses adopting the same slit width. 
This procedure corrects for slit losses since most of our LAEs were confirmed 
with HST images to be unresolved in our ground-based conditions.
The cross-calibration between the H- and K-band was independently checked
and confirmed with bright stars ($K_{Vega}=15.5$--$16.5$) included on each mask.
Some Lyman-break galaxies (LBGs) that were also placed on the masks for 
a comparison sample are more extended than LAEs in the HST image, 
and their slit losses would not be fully corrected for with the above method. 
We quantified the potential additional slit losses for the LBGs by using 
an appropriately smoothed HST image using the seeing and slit position/angles 
appropriate for the observations. We find for the small subset of LBG targets,
the additional flux losses would be smaller than 25\%. This is minimal and 
does not affect our conclusions.

\begin{deluxetable}{lclcccc}[t]
\tablecaption{MOSFIRE Near-Infrared Spectroscopy of the LACES Sample
\label{tbl:summary_observations}}
\tablewidth{0.99\columnwidth}
\tabletypesize{\scriptsize}
\tablehead{
\colhead{Name} &
\colhead{Band} &
\colhead{Date}  &
\colhead{Seeing} &
\colhead{Exp.} &
\colhead{$N^o$} &
\colhead{Ref.}
\\
\colhead{} &
\colhead{} &
\colhead{} &
\colhead{} &
\colhead{(hrs)} &
\colhead{(1)} &
\colhead{(2)}
} 
\startdata
mask1 & K & 2015 Jun 20 & $0\farcs4$--$0\farcs5$ & 3.0 & 17 & (a) \\
 & H & 2015 Jun 21 & $0\farcs4$--$0\farcs5$ & 2.5 & 17 & (a) \\
mask2 & K & 2017 Jul  31 & $0\farcs6$--$0\farcs9$ & 3.0 & 21 & (b) \\
 & H & 2017 Aug 1 & $0\farcs5$--$0\farcs9$ & 3.1 & 21 & (b) \\
mask3 & K & 2017 Aug 1 & $0\farcs5$--$0\farcs8$ & 2.3 & 17 & (b) \\
mask4 & K & 2017 Aug 1 & $0\farcs3$--$0\farcs5$ & 2.0 & 19 & (b) \\
mask5 & H & 2018 Aug 3, 4 & $0\farcs4$--$0\farcs8$ & 4.6 & 21 & (c) \\
\hline
Full $^{(\dag)}$ & K &  &  & 2.0--6.0 & 53 & (c) \\
 & H &  &  & 2.5--10.2 & 38 & (c) \\
\enddata
\tablecomments{%
(1) Number of targeted LACES sources.
(2) Relevant campaigns (a) \citet{nakajima2016};
(b) \citetalias{fletcher2019};
(c) This work.
($\dag$) Full sample in K and H taking into account multiply-observed spectra. 
}
\end{deluxetable}

The resulting K-band observations span four different masks,  including $18$ objects that were observed on
more than one mask. For each of these multiply-observed sources, we combined flux-calibrated 2D spectra
from different masks to generate a final 2D spectrum after the spatial zero points 
were aligned. Our final K-band spectroscopic sample contains $53$ LACES sources each 
with a total integration time ranging from 2.0 to 6.0 hrs. Similarly, we have H-band coverage of
$38$ LACES sources with integration times ranging from $2.5$--$10.2$ hrs. All H-band
sources have K-band coverage. 
We experimented with coaddition of the various integrations
using both 2D spectra from which 1D spectra were subsequently extracted, as well as
summation of 1D spectra individually extracted; no significant difference in S/N was found.

One dimensional (1D) spectra were produced via the summation of 5--9 pixels along the spatial direction 
centered on the expected spatial position. This width was chosen to maximize the  
signal-to-noise (S/N) ratio and corresponds approximately to twice the average seeing for
the observations.

\begin{figure*}[t]
  \centering
    \begin{tabular}{c}
      \begin{minipage}{0.42\hsize}
        \begin{center}
          \includegraphics[height=55mm]{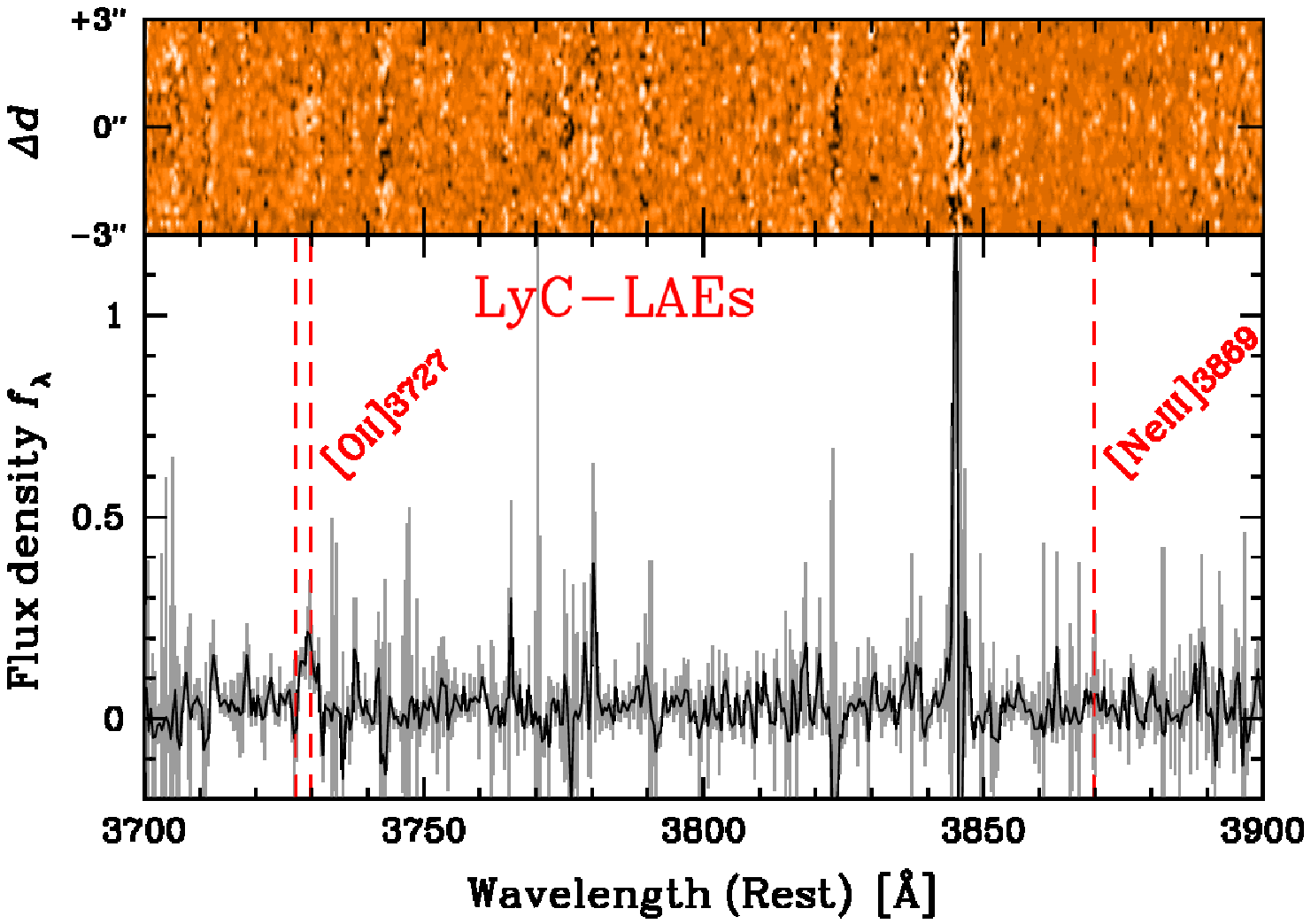}
        \end{center}     
      \end{minipage}
      \begin{minipage}{0.54\hsize}
        \begin{center}
          \includegraphics[height=55mm]{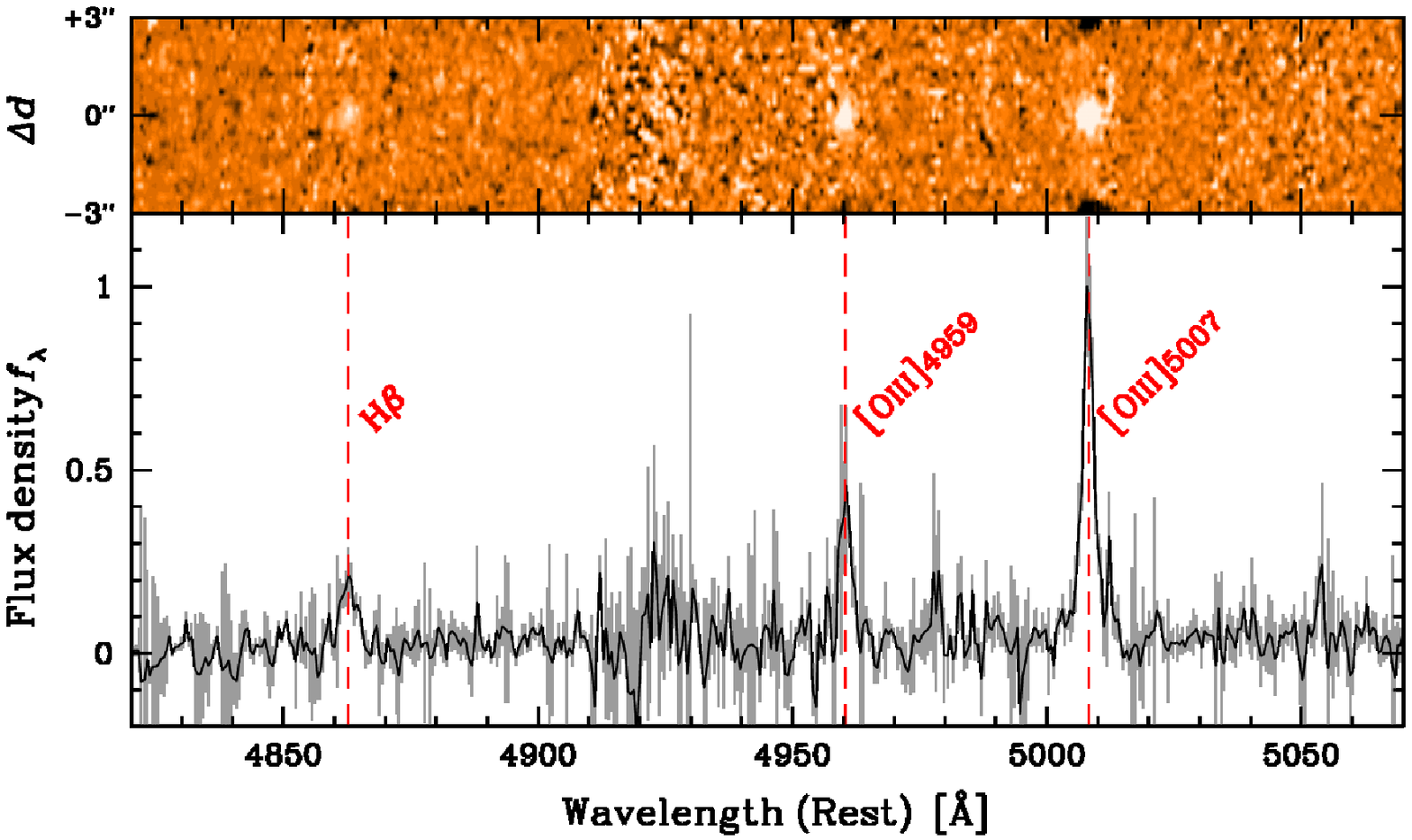}
        \end{center}
      \end{minipage}
      \\
      \begin{minipage}{0.42\hsize}
        \begin{center}
          \includegraphics[height=55mm]{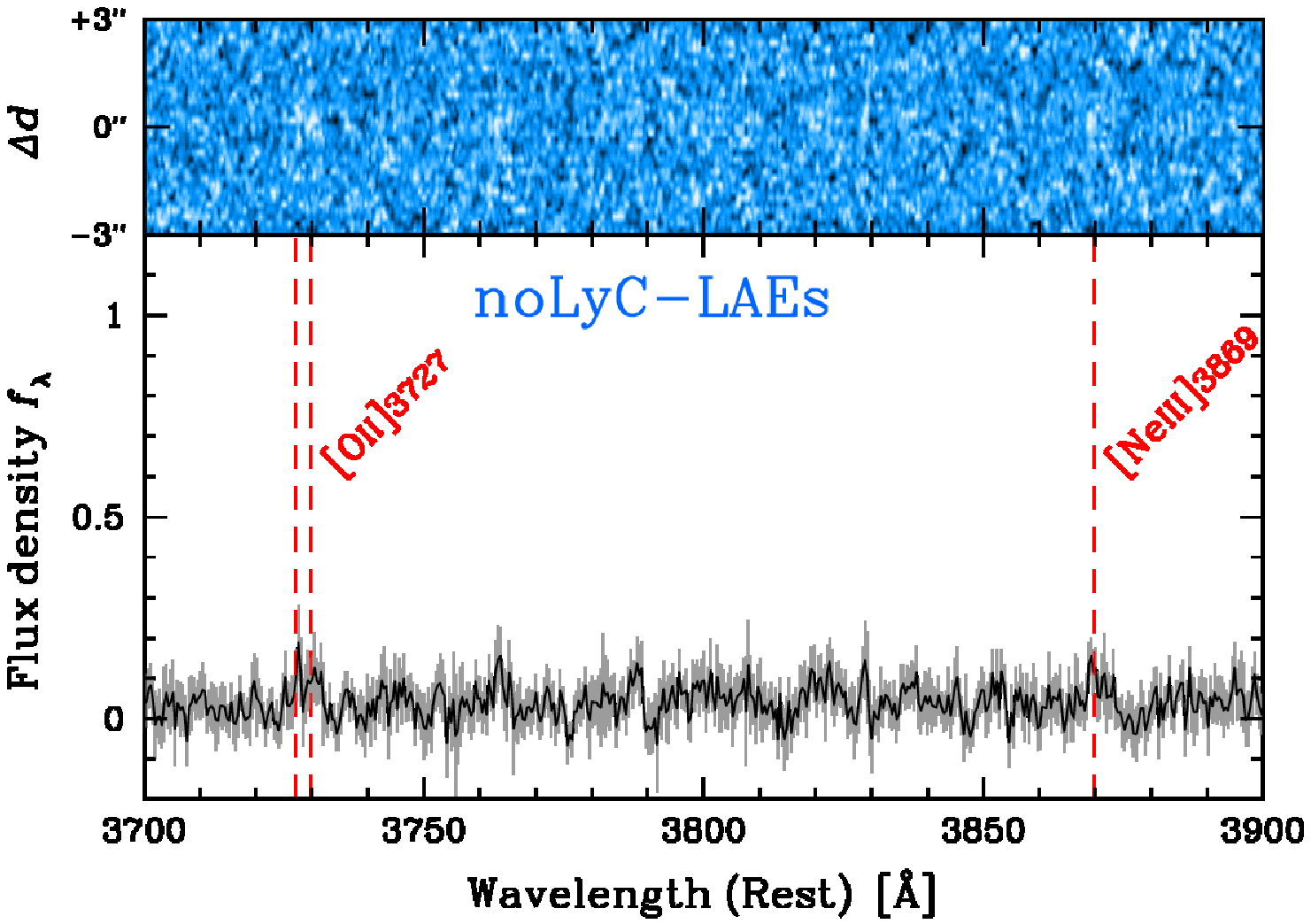}
        \end{center}     
      \end{minipage}
      \begin{minipage}{0.54\hsize}
        \begin{center}
          \includegraphics[height=55mm]{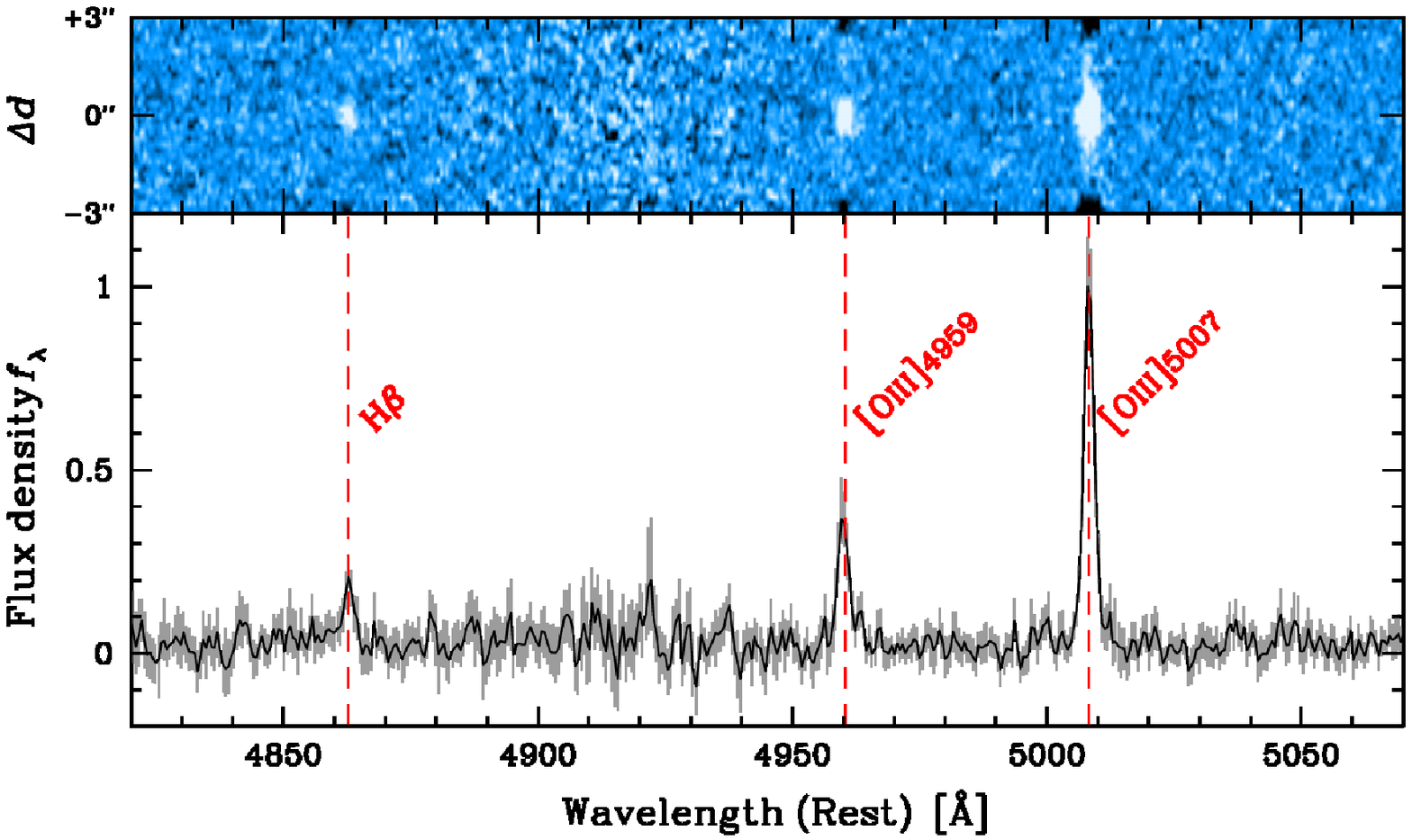}
        \end{center}
      \end{minipage}
      \\
      \begin{minipage}{0.42\hsize}
        \begin{center}
          \includegraphics[height=55mm]{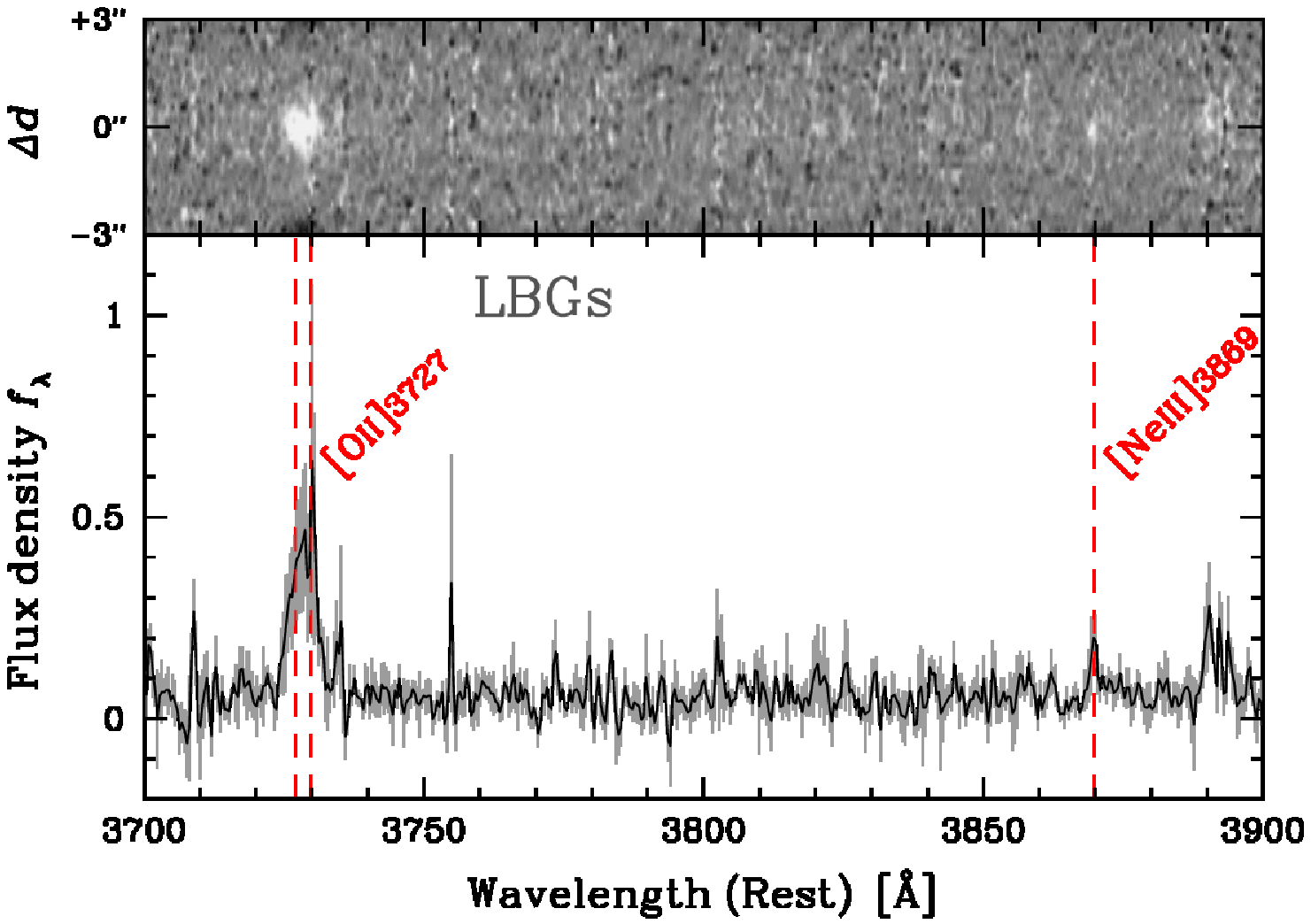}
        \end{center}     
      \end{minipage}
      \begin{minipage}{0.54\hsize}
        \begin{center}
          \includegraphics[height=55mm]{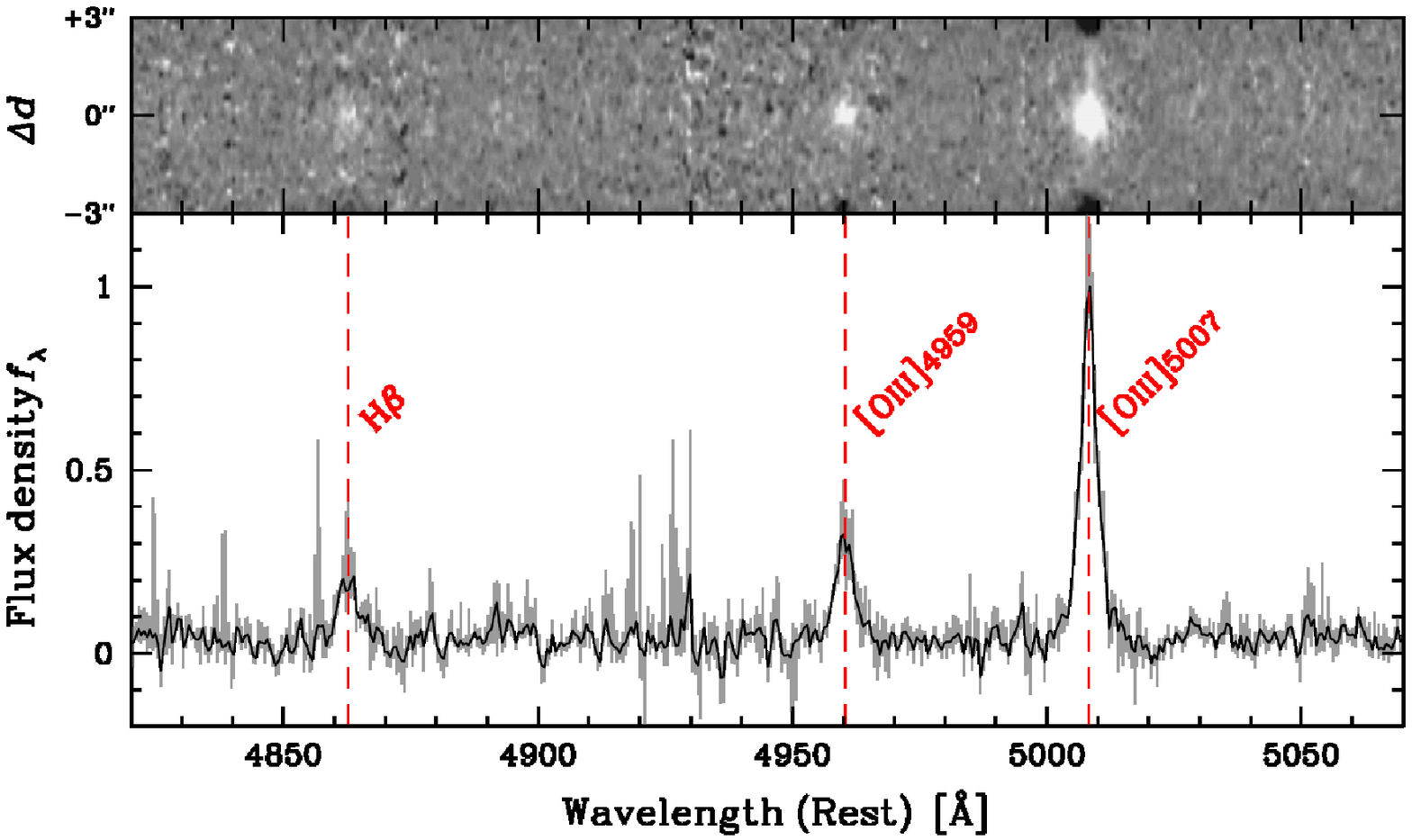}
        \end{center}
      \end{minipage}
    \end{tabular}
    \caption{%
		Composite rest-frame optical spectra of 
		LAEs with a significant HST LyC detection (LyC-LAEs; top), 
		LAEs with no detectable LyC flux (noLyC-LAEs; middle), 
		and LBGs (all undetected in LyC; bottom). 
		These spectra were generated with the \OIII-normalized individual
		spectra for measuring line flux ratios (see text for more details). 
		The grey shaded region around each spectrum refers to 
		the standard deviation of the flux density at each wavelength 
		estimated by bootstrap resampling (see the text for details). 
		The wavelengths of key diagnostic emission lines are marked 
		with a red dashed line.
		\label{fig:mosfire_composite_spectra}
    }
\end{figure*}

\subsection{Emission line identifications} \label{ssec:data_emissionlines}

Out of the $53$ K-band sources, $38$ have confirmed \Lya\ emission from our earlier optical
campaigns (see \citet{nakajima2018_z3laes} for details). For the other LAEs, their redshifts are 
fairly well-constrained from the Subaru narrowband filter used for the selection.
Using these redshifts as an initial guess, we visually examined the 1D and 2D spectra 
for detectable \OIII$\lambda\lambda 5007,4959$ and \Hb\ emission.
One or both of \OIII\ and \Hb\ were detected in the MOSFIRE K-band data for $43$ of the $53$ K-band sources
and their line fluxes were measured. We then proceeded to measure fluxes for the \OII\ 
doublet\footnote{We use the notation \OII$\lambda3727$, or simply \OII, as the sum of the doublet.
In the fitting process, we adopted two Gaussians unless otherwise noted.}
and \NeIII\ emission in the cases where H-band spectra are available ($31$ out of the $43$ with
line emission in the K-band). All H-band line fluxes were measured by fitting a Gaussian profile 
using the IRAF task \textsc{specfit} adopting the redshift and FWHM of the \OIII$\lambda 5007$. 
A constant continuum was also considered for each of the \OIII$+$\Hb\ and \OII+\NeIII\ lines
in accounting for background residuals.

To estimate the sky noise level and hence the flux uncertainties, we used more than
1000 apertures with a size equal to that adopted for the flux measurements 
spread randomly around the emission lines in the 2D spectrum after masking pixels 
heavily contaminated by OH lines. We then derived the $1\sigma$ fluctuation for each of the 
lines according to the distribution of the photon counts measured with the randomly 
distributed apertures. 
Table \ref{tbl:catalog_mosfire} lists the measured fluxes and their $1\sigma$ errors 
for the $43$ identified objects. 
Among these identified sources, there are $26$, $12$, and $12$ objects whose \Hb,
\OII, and \NeIII\ can be individually detected, respectively.

For the $10$ remaining sources with MOSFIRE spectra, three have a spectroscopic 
redshift based on \Lya, where \OIII$\lambda 5007$ cannot reliably be detected due to a strong 
OH line \footnote{This assumes the velocity offset of \Lya\ is smaller than $\sim 200$\,\kms\
corresponding to twice the resolution of MOSFIRE in the K-band as is typical for LAEs (e.g. \citealt{nakajima2018_z3laes}).}. 
For the other seven targets, without a redshift we cannot determine the expected
wavelength of \OIII$\lambda 5007$ or any other lines and hence upper limits on their fluxes.
We therefore exclude these $10$ sources in the following discussion.

\citetalias{fletcher2019} presented $12$ individual LAEs with prominent escape fractions;
\fesc\ $\sim 15-60$\%. Out of these $12$ \fesc-detected sources
\footnote{This subsample includes both the Gold and Silver classifications of \citetalias{fletcher2019}.}, 
$11$ ($9$) have  K$+$H (only $K$) band MOSFIRE spectra from which $8$ present one or more 
rest-frame optical emission lines as listed in Table \ref{tbl:catalog_mosfire}. 
These $8$ sources also have \Lya\ detections. 
The remaining $4$ prominent leakers have neither \Lya\ nor rest-frame optical emission lines;
that is they lie in the subsample of $10$ sources discussed above and will not be considered
further in this paper.

\begin{figure*}[t]
  \centering
    \begin{tabular}{cc}
      \begin{minipage}[b]{0.45\hsize}
        \begin{center}
          \includegraphics[height=55mm]{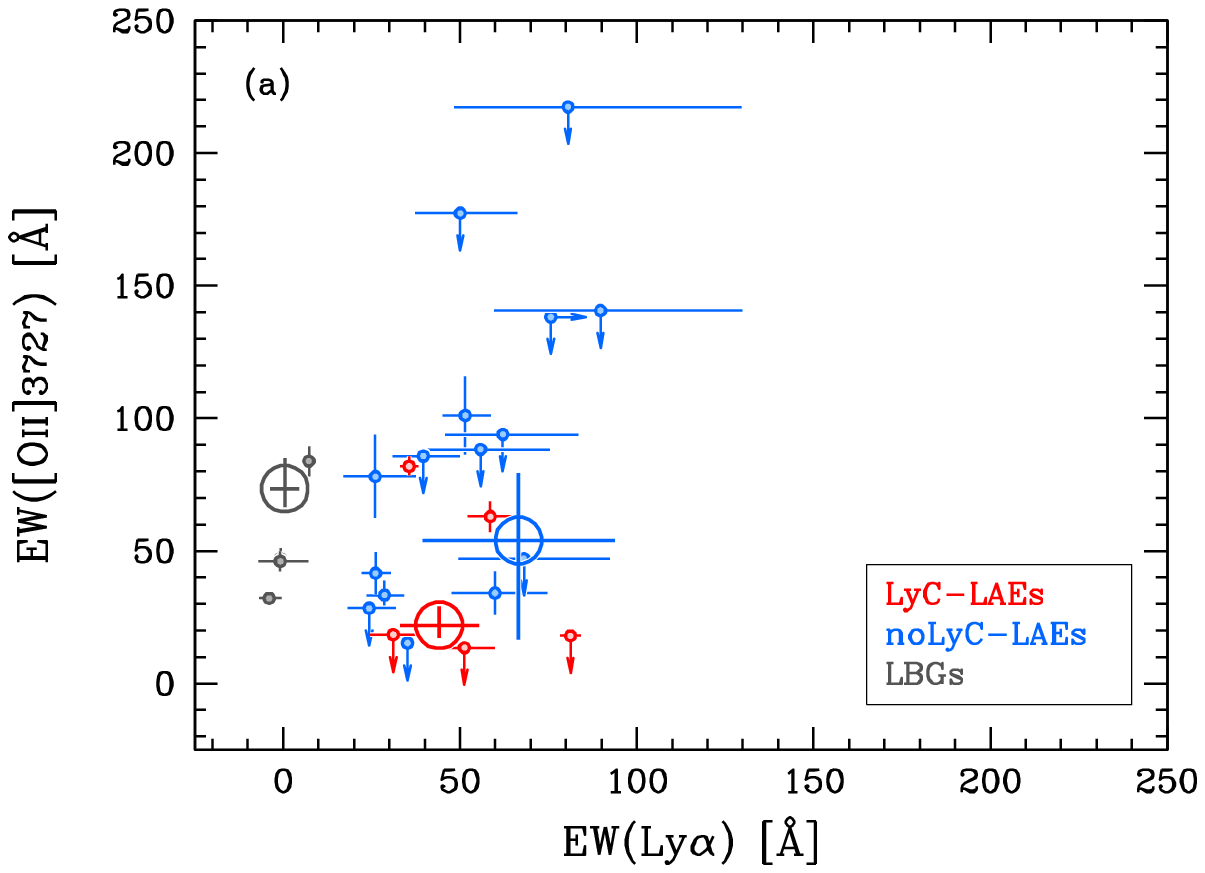}
        \end{center}     
      \end{minipage}
      \begin{minipage}[b]{0.45\hsize}
        \begin{center}
          \includegraphics[height=55mm]{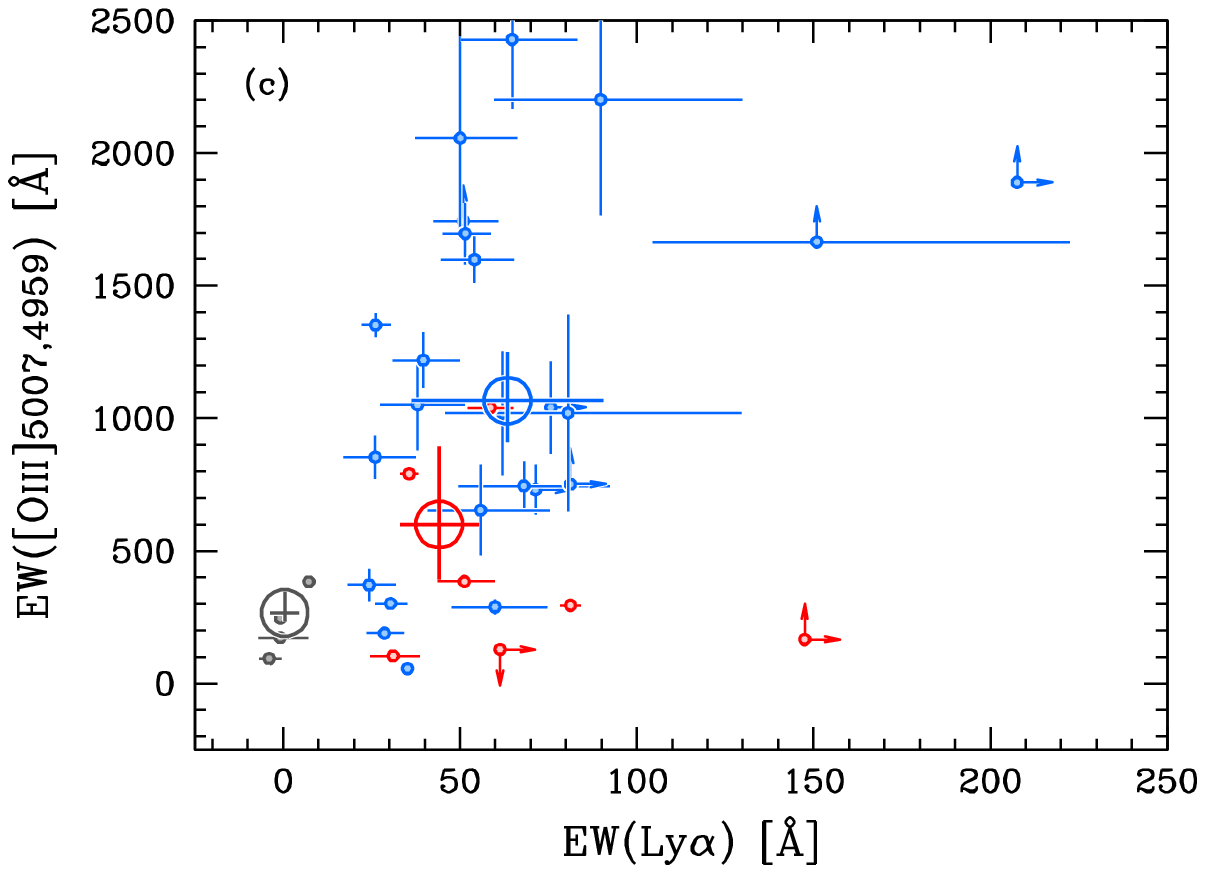}
        \end{center}     
      \end{minipage}
      \\
      \begin{minipage}[b]{0.45\hsize}
        \begin{center}
          \includegraphics[height=55mm]{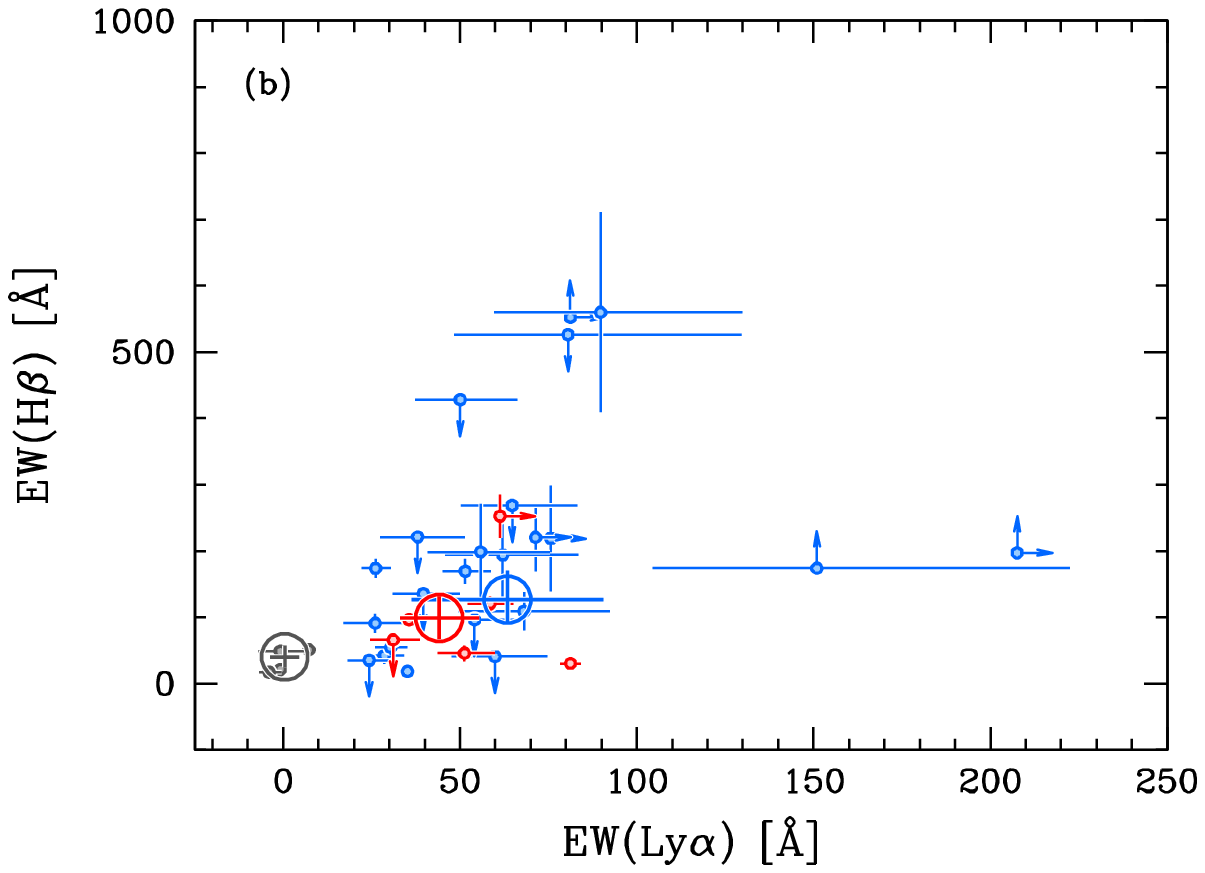}
        \end{center}     
      \end{minipage}
      \begin{minipage}[b]{0.45\hsize}
        \begin{center}
          \includegraphics[height=55mm]{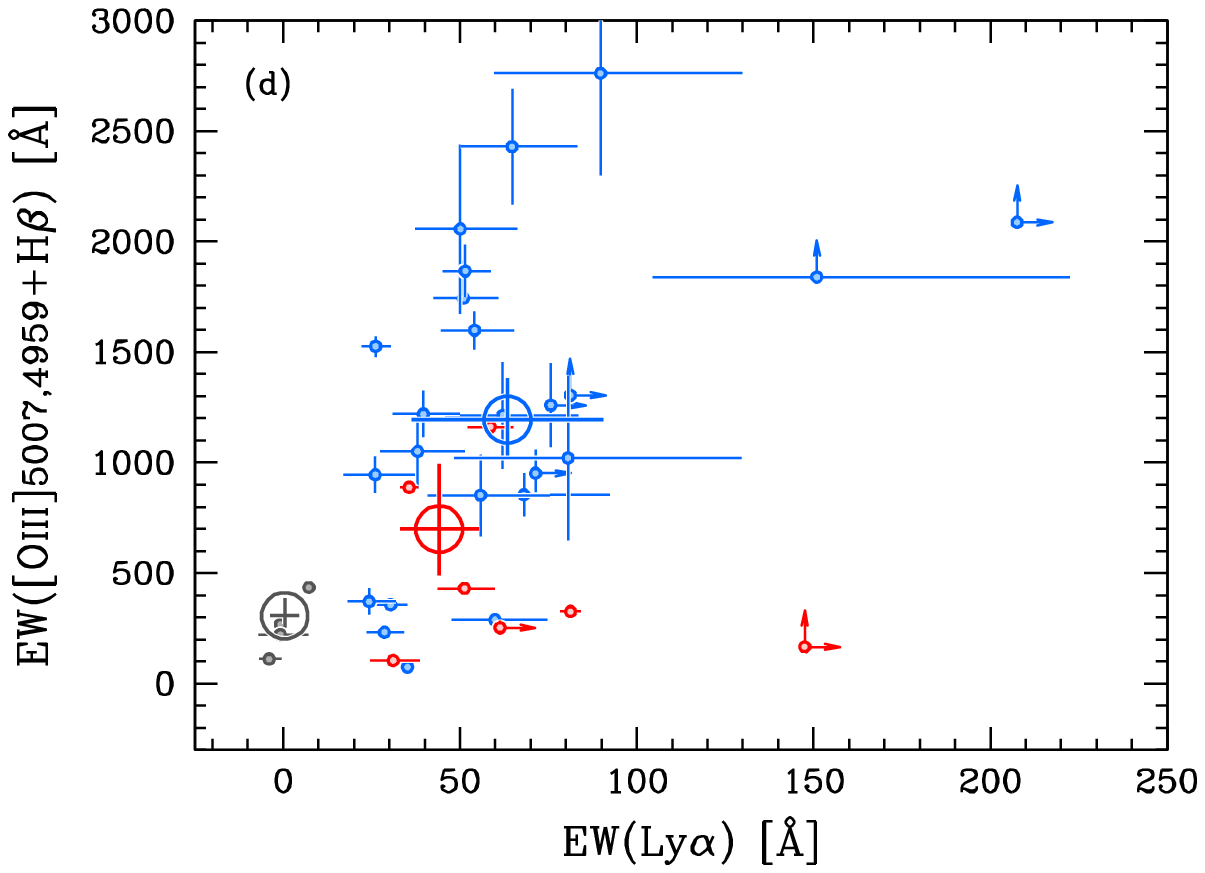}
        \end{center}     
      \end{minipage}
    \end{tabular}
    \caption{%
    		Rest-frame equivalent widths (EWs) of (a) \OII, (b) \Hb, (c) \OIII\ and (d) \OIII$+$\Hb\
		for the three LACES sub-samples (LyC-LAEs in red, noLyC-LAEs in blue, and LBGs in gray)
		as a function of EW(\Lya).
		Large symbols represent results from stacked spectra whereas the small
		faint-colored circles shows individual measurements with 3$\sigma$ upper limits shown as arrows.
		\label{fig:ews_ewlya}
    }
\end{figure*}

\begin{deluxetable}{@{}lccccc@{}}[t]
\tablecaption{Optical Emission Line Fluxes
\label{tbl:catalog_mosfire}}
\tablewidth{0.99\columnwidth}
\tabletypesize{\scriptsize}
\tablehead{
\colhead{Obj.} &
\colhead{\OII} &
\colhead{\NeIII} &
\colhead{\Hb} &
\colhead{\OIII} &
\colhead{\OIII}
\\
 &
{\tiny doublet} &
{\tiny $\lambda 3869$} &
 &
{\tiny $\lambda 4959$} &
{\tiny $\lambda 5007$}
}
\startdata
M38
& $8.5\pm 0.3$
& $1.9\pm 0.2$
& $5.3\pm 0.1$
& $6.8\pm 0.1$
& $20.5\pm 0.1$
\\
2132
& $7.6\pm 0.5$
& $1.4\pm 0.2$
& $2.8\pm 0.3$
& $5.9\pm 0.3$
& $15.3\pm 0.3$
\\
104037$^{(\dag\,S)}$
& $3.6\pm 0.1$
& $1.4\pm 0.1$
& $2.6\pm 0.1$
& $6.9\pm 0.1$
& $14.5\pm 0.1$
\\
93564$^{(\dag\,G)}$
& $1.8\pm 0.2$
& $0.4\pm 0.1$
& $2.1\pm 0.2$
& $5.2\pm 0.3$
& $13.2\pm 0.3$
\\
104511
& $0.9\pm 0.2$
& $0.9\pm 0.1$
& $2.2\pm 0.2$
& $5.2\pm 0.1$
& $12.0\pm 0.1$
\\
108679
& --
& --
& $<0.8$
& $3.5\pm 0.3$
& $10.6\pm 0.3$
\\
96688
& $4.1\pm 0.3$
& $1.7\pm 0.2$
& $1.1\pm 0.2$
& $3.1\pm 0.2$
& $10.0\pm 0.2$
\\
99330
& $1.0\pm 0.1$
& $0.5\pm 0.1$
& $1.1\pm 0.1$
& $2.6\pm 0.1$
& $8.0\pm 0.1$
\\
109140
& --
& --
& $<0.9$
& $2.1\pm 0.2$
& $5.8\pm 0.2$
\\
86861$^{(\dag\,G)}$$^{\star}$
& $<0.7$
& $0.9\pm 0.2$
& $0.7\pm 0.0$
& $1.7\pm 0.2$
& $5.4\pm 0.2$
\\
97030
& $0.9\pm 0.2$
& $0.4\pm 0.1$
& $0.7\pm 0.1$
& $1.7\pm 0.1$
& $4.7\pm 0.1$
\\
92017
& --
& --
& $<0.8$
& $0.8\pm 0.2$
& $4.0\pm 0.2$
\\
106500
& $<0.5$
& $0.5\pm 0.1$
& $1.2\pm 0.1$
& $1.2\pm 0.2$
& $3.5\pm 0.2$
\\
104097
& $1.8\pm 0.1$
& $<0.3$
& $1.2\pm 0.1$
& $1.0\pm 0.1$
& $3.1\pm 0.1$
\\
102334
& --
& --
& $0.7\pm 0.2$
& $1.0\pm 0.2$
& $2.9\pm 0.2$
\\
94460$^{(\dag\,S)}$
& $<0.2$
& $<0.2$
& $0.4\pm 0.1$
& $0.9\pm 0.1$
& $2.8\pm 0.1$
\\
102826
& $2.0\pm 0.1$
& $<0.3$
& $0.7\pm 0.1$
& $<0.3$
& $2.8\pm 0.1$
\\
107585
& $<0.4$
& $<0.6$
& $1.0\pm 0.2$
& $0.8\pm 0.2$
& $2.3\pm 0.2$
\\
110896
& --
& --
& $<0.7$
& $<0.7$
& $2.3\pm 0.2$
\\
89114
& $<0.4$
& $<0.4$
& $<0.3$
& $0.8\pm 0.1$
& $2.3\pm 0.1$
\\
99415
& $<0.4$
& $<0.3$
& $0.5\pm 0.1$
& $0.5\pm 0.1$
& $1.9\pm 0.1$
\\
97081
& $<0.2$
& $<0.2$
& $0.3\pm 0.1$
& $0.6\pm 0.1$
& $1.9\pm 0.1$
\\
90428
& --
& --
& $<0.4$
& $<1.0$
& $1.7\pm 0.3$
\\
93474
& $<0.3$
& $<0.4$
& $<0.5$
& $0.6\pm 0.1$
& $1.7\pm 0.1$
\\
85165
& $<1.0$
& $<1.6$
& $0.8\pm 0.2$
& $0.8\pm 0.2$
& $1.6\pm 0.2$
\\
92616$^{(\dag\,G)}$
& --
& --
& $<0.3$
& $<0.4$
& $1.5\pm 0.1$
\\
104147
& $<0.2$
& $0.2\pm 0.1$
& $<0.2$
& $0.2\pm 0.1$
& $1.4\pm 0.1$
\\
92219
& $<0.3$
& $<0.3$
& $<0.3$
& $<0.2$
& $1.4\pm 0.1$
\\
93004
& --
& --
& $<0.4$
& $0.7\pm 0.1$
& $1.4\pm 0.1$
\\
92235
& $0.5\pm 0.1$
& $<0.3$
& $0.4\pm 0.1$
& $<0.4$
& $1.4\pm 0.1$
\\
97254
& $<0.2$
& $<0.2$
& $0.3\pm 0.1$
& $0.8\pm 0.1$
& $1.3\pm 0.1$
\\
97176
& $0.3\pm 0.1$
& $0.3\pm 0.1$
& $<0.2$
& $0.4\pm 0.1$
& $1.3\pm 0.1$
\\
103371
& $<0.2$
& $<0.2$
& $0.2\pm 0.1$
& $0.9\pm 0.1$
& $1.3\pm 0.1$
\\
89723
& $<0.2$
& $<0.2$
& $<0.2$
& $0.5\pm 0.1$
& $1.3\pm 0.1$
\\
110290
& $<0.3$
& $<0.3$
& $0.4\pm 0.1$
& $<0.6$
& $1.1\pm 0.2$
\\
93981
& --
& --
& $0.4\pm 0.1$
& $0.4\pm 0.1$
& $1.1\pm 0.1$
\\
105937$^{(\dag\,S)}$
& $<0.4$
& $<0.3$
& $<0.9$
& $0.4\pm 0.1$
& $1.0\pm 0.1$
\\
91055
& $<0.3$
& $<0.2$
& $0.3\pm 0.1$
& $0.3\pm 0.1$
& $0.9\pm 0.1$
\\
107677
& --
& --
& $<1.0$
& $<0.5$
& $0.9\pm 0.2$
\\
95217
& $<0.4$
& $<0.2$
& $<0.6$
& $<0.4$
& $0.8\pm 0.1$
\\
97128
& --
& --
& $0.7\pm 0.2$
& $<0.5$
& $0.7\pm 0.2$
\\
101846$^{(\dag\,S)}$
& $<0.3$
& $<0.3$
& $<0.3$
& $<0.3$
& $0.4\pm 0.1$
\\
90675$^{(\dag\,G)}$$^{\mathparagraph}$
& --
& --
& $2.0\pm 0.2$
& $<0.5$
& $<0.5$
\enddata
\tablecomments{%
Fluxes and their $1\sigma$ errors are given in units of $10^{-17}$\,erg\,s$^{-1}$\,cm$^{-2}$.
Upper-limits represent the $3\sigma$ values.
($\dag$) LyC leaking candidates from \citetalias{fletcher2019}.
The $G$ and $S$ denotes the Gold and Silver sample, respectively. 
($\star$) The single LAE-AGN in the LACES sample.
($\mathparagraph$) This  LAE is likely an extremely metal-poor galaxy 
on account of its low \OIII$/$\Hb, large \Hb\ and \Lya\ EWs, and high \xiion\ 
(see also Table \ref{tbl:physical_properties}).
}
\end{deluxetable}

\begin{deluxetable}{lccc}[t]
\tablecaption{Subsamples of the LACES MOSFIRE campaign
\label{tbl:summary_subsamples}}
\tablewidth{0.99\columnwidth}
\tabletypesize{\scriptsize}
\tablehead{
\colhead{Subsample} &
\colhead{for line ratios } &
\colhead{for EWs in K/H} &
\colhead{for \xiion}
} 
\startdata
LyC-LAEs$^{(\dag)}$ & $5$ & $5$/$5$ & $6$ \\
noLyC-LAEs & $20$ & $24$/$17$ & $29$\\
LBGs & $5$ & $5$/$5$ & $6$
\enddata
\tablecomments{%
($\dag$) A single AGN-LAE, AGN86861, whose LyC radiation was identified 
in \citetalias{fletcher2019} has been removed for the stacking analysis and is
not counted here.
}
\end{deluxetable}

\begin{figure}[h!]
	\begin{center}
		\includegraphics[width=0.99\columnwidth]{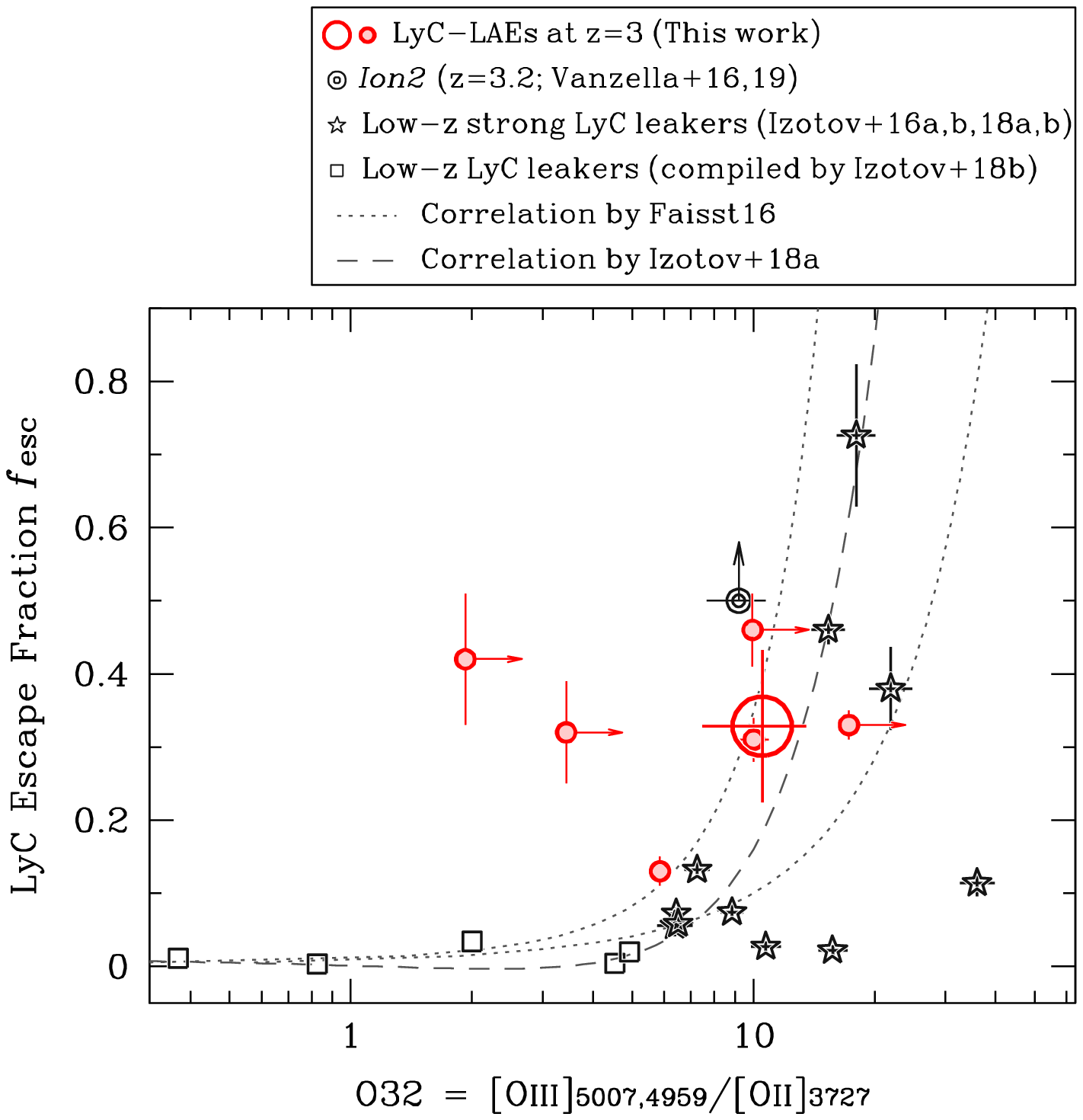}
        \end{center}
	\caption{%
		The O32 line ratio vs. escape fraction \fesc\ for the sources with a LyC detection.
		The large circle represents the LyC-LAEs subsample composite. Other symbols
		and curves represent literature measures and relations, respectively,
		as indicated by the legend. The correlation suggests that a 
		prerequisite for a high \fesc\ is a large O32 value. 
		\label{fig:o3o2_fesc}
    }
\end{figure}

\subsection{Stacked spectra} \label{ssec:data_stackedspectra}

Despite our significant integration times, we only directly detect individual \OII\ emission lines 
in a subset of our data (Section \ref{ssec:data_emissionlines}, Table \ref{tbl:catalog_mosfire}).
To exploit the full diagnostic value of the rest-frame optical emission lines, we therefore developed a 
stacking procedure for various subsamples of the LACES catalog. Our goal is to use the
stacked spectra to derive average line strengths, line ratios and measures of the ionizing radiation field
\xiion\ (see \S\ref{ssec:analysis_xiion} for definition and more details) and to correlate these properties 
with the strength of LyC leakage as determined in \citetalias{fletcher2019}.

Accordingly, we divided our spectroscopic sample into three subsamples:
LAEs with a clear LyC detection defined as a $>4\sigma$ detection in \citetalias{fletcher2019}
(hereafter called ``LyC-LAEs'' subsample), those LAEs without a clear LyC signal 
(``noLyC-LAEs'' subsample), and LBGs, none of which
reveals a LyC signal (``LBGs'' subsample). To distinguish LAEs from
LBGs we adopted a rest-frame equivalent width (EW) of $20$\,\AA, derived spectroscopically
and/or photometrically, as the demarcation level. The LyC-LAEs subsample includes both the 
Gold and Silver subsamples in \citetalias{fletcher2019} but excludes the non-thermal source AGN86861.
The numbers of sources in each of the subsamples are given in 
Table \ref{tbl:summary_subsamples}.

It is important to note that the individual spectra must be normalized in a different manner 
prior to stacking depending on the physical quantity we seek to measure. 
For individual line ratios, we use the \OIII\ line flux, whereas for EWs and the
\xiion\ parameter we use the rest-frame optical and UV continuum,
respectively, derived from the HST/F160W and the Subaru optical photometry.
Naturally for line ratios, we require both H- and K-band spectra,
whereas for individual measures of \Hb\ or \OIII\ only K-band data is required.
Thus the numbers of useful spectra for stacking varies according to the
physical quantity concerned. The details are given in Table \ref{tbl:summary_subsamples}.

We adopted a stacking procedure very similar to that described in \citet{nakajima2018_z3laes}.
Briefly, using the individual flux-calibrated spectra in K (H), we shifted each to its rest-frame and 
rebinned the spectrum to a common dispersion of 0.55 (0.40)\,\AA\ per pixel.
The spectra were then median-stacked with the appropriate normalization as 
explained above. To exclude positive and negative sky subtraction residuals, we rejected 
an equal number of the highest and lowest outliers at each pixel 
corresponding in total to $\simeq 5$ percent of the data. Using
an averaging method led to spectra almost indistinguishable from using the median. 

To evaluate sample variance and statistical noise, we adopted a bootstrap technique
similar to that described in \citet{nakajima2018_z3laes}. 
We generated $1000$ fake composite spectra from the chosen sample. 
Each fake spectrum was constructed in the same way, using the same number of spectra 
as the actual composite, but with the list of input spectra formulated by selecting 
spectra at random, with replacement, from the full list. 
With these $1000$ fake spectra, we derived the standard deviation at each spectral pixel. 
The standard deviations are taken into account in calculating the uncertainties of each line flux.

The composite spectra for the three subsamples normalized by their \OIII\ fluxes are shown in 
Figure \ref{fig:mosfire_composite_spectra}. It is evident that all the key diagnostic emission lines are 
significantly identified. The difference between the stacked spectra of LAEs and LBGs is immediately
apparent e.g. in the \OIII$/$\OII\ and \NeIII$/$\OII\ line ratios.

\begin{figure*}[t]
	\begin{center}
		\includegraphics[width=0.95\textwidth]{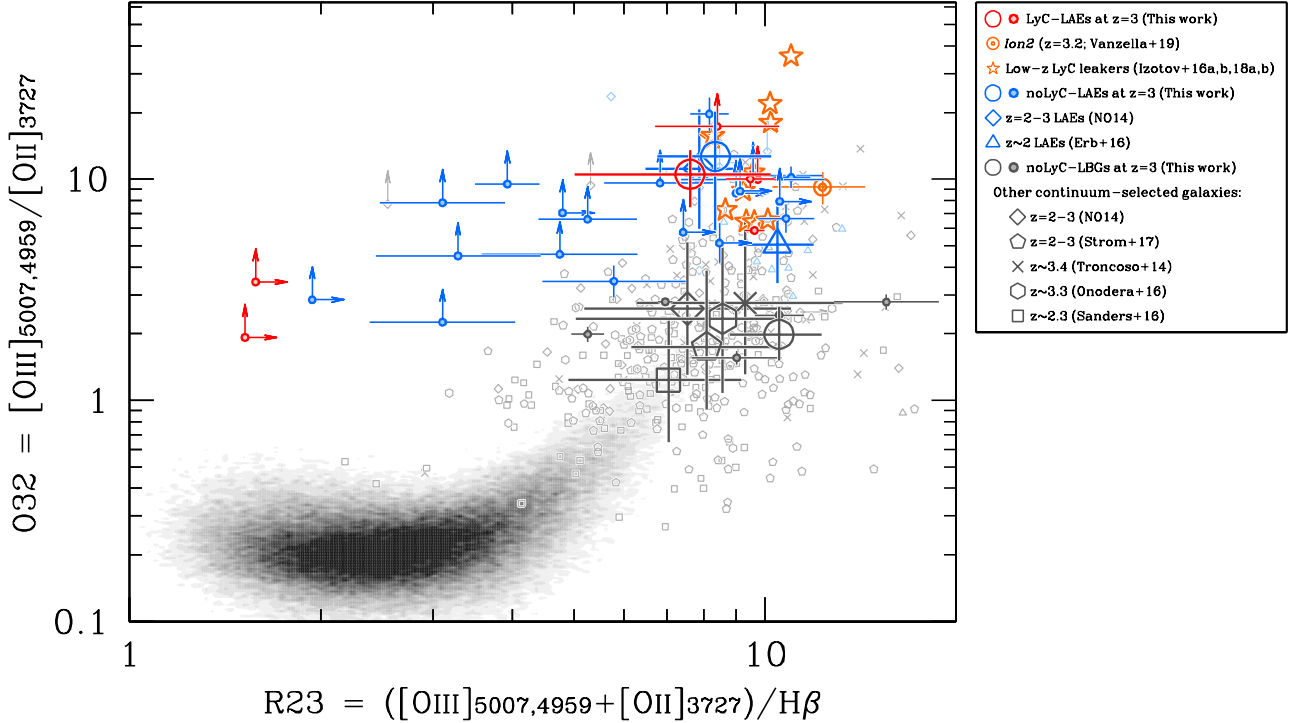}
        \end{center}
	\caption{%
		O32 vs. R23 diagram for the LACES and other samples.
		The LACES subsamples are shown with the same symbols and the colors
		as in Figure \ref{fig:ews_ewlya}.
		Orange symbols show known LyC leakers, and blue and grey
		symbols high-$z$ LAEs and continuum-selected galaxies, respectively, 
		compiled from the literature as shown by the legend. 
		If O32 is provided without a \OIII$\lambda 4959$ contribution, we
		correct for it assuming the theoretical \OIII$\lambda5007/4959$ line ratio
		of $2.98$ \citep{SZ2000}.
		Arrows provide $3\sigma$ lower limits. 
		Grey shading illustrates the equivalent distribution for nearby SDSS galaxies. 
		\label{fig:o3o2r23}
    }
\end{figure*}

\subsection{Dust correction to the nebular spectra} \label{ssec:data_dust_correction}

Prior to quantitative analysis, it is necessary to consider corrections for dust reddening, 
particularly for line flux ratios and the \xiion\ parameter. Since multiple Balmer emission lines 
cannot be reliably identified in the individual spectra, the amount of reddening must be
estimated using the stellar continuum, assuming that nebular emission 
and the stellar continuum suffer similar attenuation. 
Although this assumption remains open to debate at high-$z$, 
it appears reasonable for young, low-mass star-forming galaxies appropriate for our sample
(SFR $\sim 1$--$10$\Msun\,yr$^{-1}$ and M$_{\star} \sim 10^{8.5}$--$10^{9.5}$\Msun; 
e.g. \citealt{reddy2015}).

Earlier studies have tended to indicate LAEs are largely dust-free systems
(e.g. \citealt{erb2016,trainor2016}).
Using the SMC extinction curve \citep{gordon2003} and 
the BPASS SEDs, \citetalias{fletcher2019} conducted SED model fitting to constrain the stellar 
population parameters as well as the amount of dust attenuating the stellar continuum emission. 
That analysis returned an almost negligible dust attenuation for LAEs 
\textit{irrespective of LyC identification} with \ebv\ $\simeq 0.01$. 
Such a small amount of dust is also discussed and supported by our pilot observations in 
\citet{nakajima2016}, where small Balmer decrements for two bright LAEs were shown to
be consistent with zero reddening.  
Furthermore, \citet{tang2019} illustrate a monotonic decrease of nebular attenuation 
with increasing EW of \OIII, showing that the most extreme line emitters with 
EW(\OIII) $\gtrsim 800$\,\AA\ have almost no dust attenuation effect on the nebular emission lines.
The relationship derived in \citet{tang2019} supports the assumption of little dust correction
for the LAE sample, given their extremely strong \OIII\ emission in general 
(\S\ref{sec:analysis}).
A similar implication is also drawn in \citet{erb2016} using the O32 line ratio.
A larger value of \ebv\ $\simeq 0.10$ was inferred on 
average for the LBG subsample following the same SED fitting procedure.

Because the \ebv\ value is generally uncertain for individual faint sources, 
we adopt the average of \ebv\ $=0.01$ for all the individual and composite spectra 
for the LAE subsamples, and \ebv\ $=0.10$ for the LBGs subsample 
in the following analysis.

\section{Analysis} \label{sec:analysis}

\subsection{Emission lines as a function of \fesc} \label{ssec:analysis_emissionlines}

We now discuss the correlation between the LyC detections presented in \citetalias{fletcher2019}
and both the individual and stacked line measurements derived for the various subsamples 
of our MOSFIRE spectra. We begin with individual line measures updating and extending 
some of the results presented in \citetalias{fletcher2019}.

Figure \ref{fig:ews_ewlya} shows that LAEs on average present an intense \OIII\ 
emission line with a rest-frame EW of $\simeq 600-1100$\,\AA\ , 
consistent with the results of our pilot MOSFIRE program \citep{nakajima2016}. 
Our enlarged spectroscopic data also reveals more intense \Hb\ emission with an EW
of $>100$\,\AA\ . A combined EW of \OIII$+$\Hb\ of $\simeq 700-1200$\,\AA\ confirms
the suggestion that such intermediate redshift LAEs are close analogs of galaxies 
in the reionization era where the similarly large EWs have been inferred from 
Spitzer photometry
(e.g. \citealt{smit2015,roberts-borsani2016}; see also \citealt{tang2019,reddy2018_mosdef}).

One of the most interesting questions we can now consider is, via our various spectroscopic
measures, what is the physical origin of the bimodal nature of LyC emission seen in the 
LACES sample \citepalias{fletcher2019}. 
In \citetalias{fletcher2019}, we presented a preliminary EW(\OIII) distribution that
revealed no significant difference between those LAEs with and without a LyC detection. 
We can see this is also the case in Figure \ref{fig:ews_ewlya} and the conclusion would 
not be changed after correcting by a ($1-$\fesc) factor in order to compensate for 
escaping (i.e. unconsumed) numbers of ionizing photons.

However, when we turn to consideration of the \OIII$/$\OII\ ratio which we could not
consider in \citetalias{fletcher2019}, a more interesting result emerges. 
This ratio represents the degree of ionization in the hot ISM and, using photoionization models, 
\citet{NO2014} argued that intense high ionization lines, e.g. \OIII, and weaker low ionization lines, 
e.g. \OII, could arise from density-bounded \HII\ regions. The associated porosity of the star-forming 
regions to ionizing radiation would lead to a high \fesc\ 
(see also \citealt{JO2013,zackrisson2013,behrens2014}).

Figure \ref{fig:o3o2_fesc} presents the relationship between \fesc\ and \OIII$/$\OII\ line ratio
for the LACES LyC-LAEs subsample. Our stacked LyC subsample with an average escape 
fraction \fesc\ $\sim 0.35$ has a large \OIII$/$\OII\ line ratio of $\simeq 10$. Combining
this measurement with individual LyC leaking sources
at low-$z$ \citep{izotov2016_nature,izotov2016_4more,izotov2018_46per,izotov2018_5more}
as well as a single $z=3$ LyC emitter, \textit{Ion2} \citep{vanzella2016_ion2,debarros2016,vanzella2019_ion2},
strengthens the positive correlation presented by \citet{izotov2018_46per} and \citet{faisst2016}. 
{\it Figure \ref{fig:o3o2_fesc} shows that a large \OIII$/$\OII\ ratio 
is a necessary condition for sources with a high \fesc. Significantly, the high escape
fraction inferred (\fesc\ $>0.1$) is approximately the lower limit necessary if
star-forming galaxies govern the reionization process
(e.g. \citealt{robertson2015}). In our sample, these are only found if the \OIII$/$\OII\ ratio 
exceeds $\sim 6-7$.}

On the other hand, a large \OIII$/$\OII\ line ratio need not in every case imply a prominent 
LyC flux as can be inferred also from the composite spectrum of the noLyC-LAEs 
(middle panel in Figure \ref{fig:mosfire_composite_spectra}). 
This contradiction is also apparent in low-redshift green pea galaxies 
\citep{izotov2018_5more,jaskot2019}. 
We evaluate this further in Figure \ref{fig:o3o2r23}, where we compare our LAEs with and 
without a LyC detection in the \OIII$/$\OII\ line ratio versus R23-index diagnostic diagram.
This diagram is widely used to examine the gas-phase metallicity and ionization state
in the local universe (e.g. \citealt{KD2002}) as well as at $z=2-4$ 
(e.g. \citealt{maiolino2008,NO2014,shapley2015,onodera2016,strom2017,sanders2019}).
A relatively large scatter in the R23-index at fixed \OIII$/$\OII\ is seen although for
those sources the \OIII$/$\OII\ measure is only a lower limit (Table \ref{tbl:catalog_mosfire}).
This may reflect a low metallicity tail to the distribution for which much larger \OIII$/$\OII\  indices
are implied.
Following \citet{NO2014}, \citet{izotov2016_nature,izotov2016_4more}, and 
\citet{nakajima2016}, we can argue that LAEs and low-$z$ LyC-confirmed green pea galaxies 
share the similarity in the line emission properties. 
This work can additionally deduce that both LyC-detected and non-detected LAEs 
share similar high \OIII$/$\OII\ line ratios (see also \citealt{erb2016}). 
Such large ratios, indicative of a high ionization parameter are not characteristic of 
continuum-selected sample at a similar redshift 
\citep{troncoso2014,onodera2016,sanders2016_density,strom2017} as is confirmed by 
our own LBG subsample.

\begin{figure}[t]
	\begin{center}
		\includegraphics[width=0.95\columnwidth]{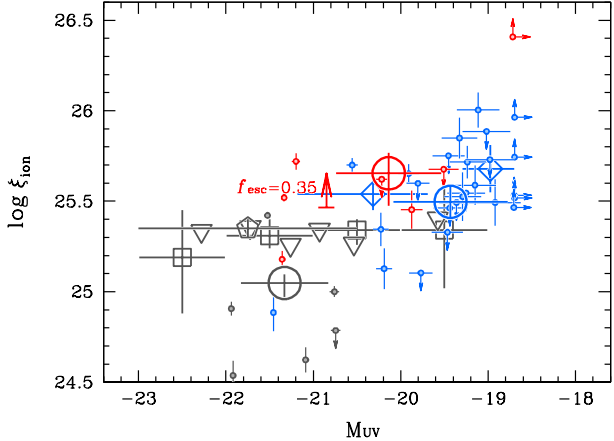}
        \end{center}
	\caption{%
		Ionizing photon production efficiency \xiion\ 
		as a function of UV absolute magnitude. 
		The symbols and colors for the 
		LACES subsamples are as shown in Figure \ref{fig:ews_ewlya}.
		Large symbols representing average \xiion\ values are derived
		from the stacked spectra (Section \ref{ssec:data_stackedspectra}).
		The individual and the stacked \xiion\ are all dust-corrected
		as detailed in Section \ref{ssec:data_dust_correction}.	
		The red upward-arrow indicates the average degree of correction 
		from \xiionzero\ to \xiion
		for the LyC-LAEs.
		Blue open diamonds present \xiion\ measurements
		for LAEs at $z=3$ \citep{nakajima2018_z3laes}, and 
		grey open symbols refer to \xiion\ measurements
		for continuum-selected galaxies at $z\simeq 2-4$
		(squares from \citealt{bouwens2016_xiion}, inverse triangles from
		\citealt{shivaei2018}, pentagon from \citealt{nakajima2018_vuds}).
		\label{fig:xi_Muv}
    }
\end{figure}

\begin{figure*}[t]
  \centering
    \begin{tabular}{c}
      \begin{minipage}{0.49\hsize}
        \begin{center}
          \includegraphics[height=62.5mm]{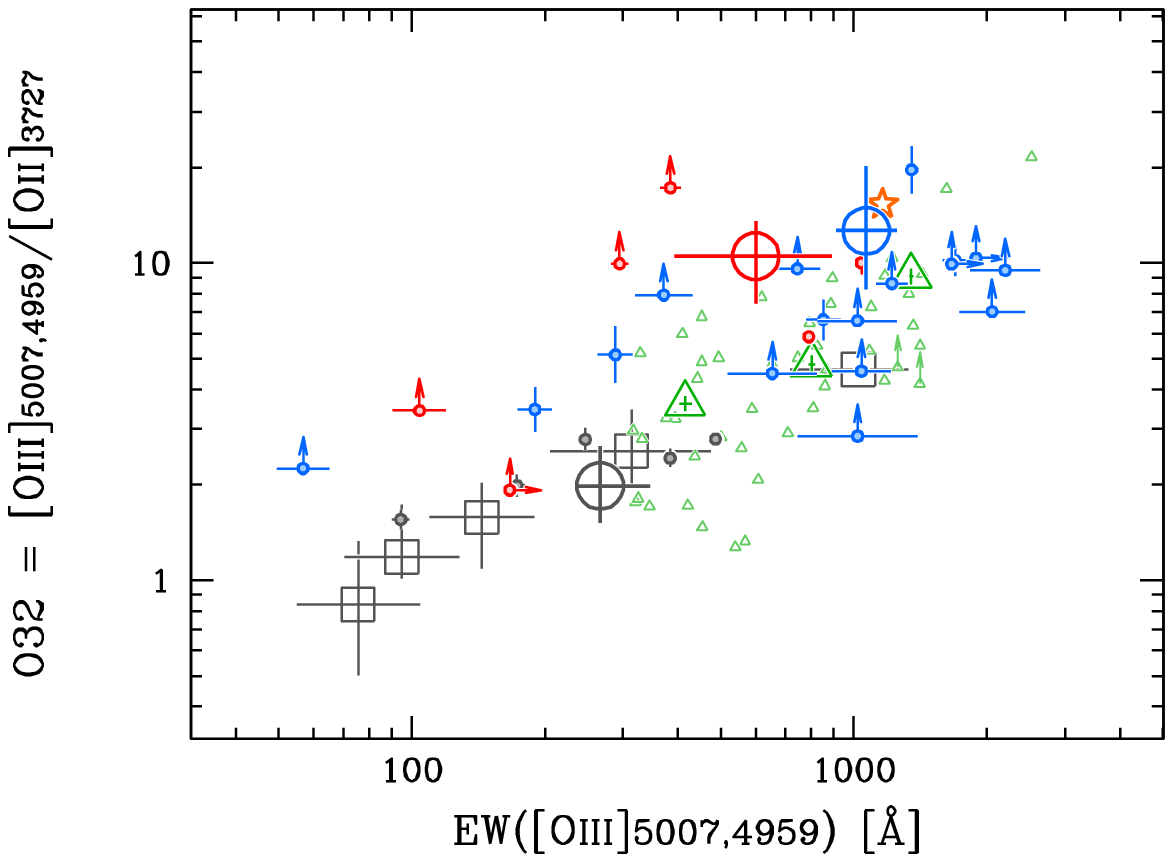}
        \end{center}     
      \end{minipage}
      \begin{minipage}{0.49\hsize}
        \begin{center}
          \includegraphics[height=62.5mm]{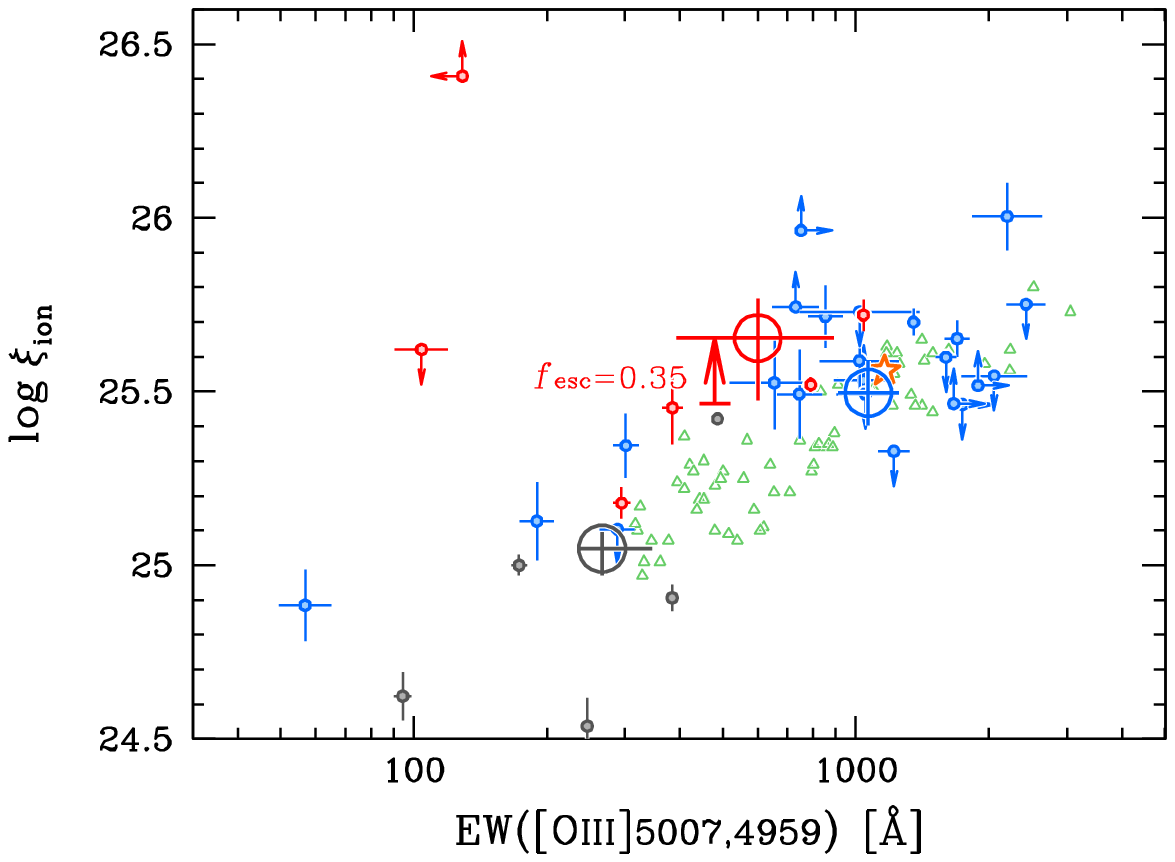}
        \end{center}
      \end{minipage}
    \end{tabular}
    \caption{%
     		Relationship of \OIII$/$\OII\ ratio (O32; Left) and ionizing photon production 
		efficiency (\xiion; Right) as a function of EW(\OIII).
		The LACES sources are plotted with circles as shown in Figure \ref{fig:ews_ewlya}.
		Green triangles present individual (small) and composite (large) measurements of 
		extreme emission line galaxies (EELGs) at $z=1.3-2.4$ 
		\citep{tang2019}. 
		Orange star illustrates a low-$z$ strong LyC emitting galaxy, J1154+2334,
		with \fesc\ $=0.46$ \citep{izotov2018_46per,schaerer2018}.
		In the left panel, grey open squares show the O32 vs. EW(\OIII) relationship derived 
		with the composites of $z\sim 2.3$ continuum-selected galaxies from MOSDEF
		\citep{sanders2019}. 
		If O32 and/or EW(\OIII) is provided without a \OIII$\lambda 4959$ contribution, 
		we correct for it assuming the theoretical \OIII$\lambda5007/4959$ line ratio
		of $2.98$ \citep{SZ2000}.
		\label{fig:o3o2_xi_ewo3}
    }
\end{figure*}

\subsection{Ionizing Radiation Field} \label{ssec:analysis_xiion}

We finally consider the hardness of the ionizing radiation field which is a further quantity related
to the escaping radiation. The efficiency of ionizing photon production is conventionally parameterized
by \xiion\ defined as: 
\begin{equation}
\xi_{\rm ion,0} = \frac{Q_{{\rm H}^0}}{L_{\rm UV}}.
\end{equation}
The number of ionizing photons, $Q_{{\rm H}^0}$, can be determined via hydrogen recombination 
lines \Hb\ (e.g \citealt{LH1995}), and the UV luminosity, $L_{\rm UV}$, is derived from the Subaru
photometry \citepalias{fletcher2019}. The subscript $0$ in \xiionzero\ indicates that the escape fraction 
of ionizing photons in this relation is assumed to be zero.
The measurable quantity \xiion\ can then be derived by dividing \xiionzero\ by $(1-f_{\rm esc})$.
Our pilot MOSFIRE program together with a rest-frame UV spectroscopic campaign conducted with
VIMOS on the VLT indicated that LAEs have \xiion\ values significantly larger than those for
continuum-selected LBGs (\citealt{nakajima2016,nakajima2018_z3laes}; see also \citealt{matthee2017}).

Figure \ref{fig:xi_Muv} provides the distribution of \xiion\ for the various LACES subsamples.
For the LyC-LAEs subsample we adopted an average escape fraction of $f_{esc}$=0.35 to make
the conversion. By improving the detectability of \Hb\ through our recent MOSFIRE campaign, 
we can confirm our earlier suggestion that \xiion\ is significantly larger for LAEs than for 
continuum-selected LBGs.
But again, we can see that both LyC-detected and non-detected LAEs subsamples have comparable values, 
$\log$\,\xiion\ $\simeq 25.5-25.7$, providing further evidence that the two populations of
LAEs are spectroscopically indistinguishable.  LAEs with LyC leakage are more efficient producers of 
ionizing photons at a given UV luminosity by $\simeq 0.3-0.4$ dex compared to continuum-selected 
LBGs but by only $\simeq 0.1$ dex with respect to our noLyC LAEs.
A similarly high \xiion\ is reported from another LyC leaker, 
\textit{Ion3} \citep{vanzella2018_ion3}.

\section{Discussion} \label{sec:discussion}

The original motivation for this series of papers was the view, following \citet{NO2014}, 
that the unusually large O32 indices of LAEs (Figure \ref{fig:o3o2r23}) implied 
density-bound star forming regions and thus a higher escape fraction of ionizing photons 
than for typical Lyman break galaxies. In this sense, therefore, we considered the population 
as valuable analogs of sources in the reionization era for which direct measures of 
LyC leakage are currently not possible.

In this paper, we have shown in Figure \ref{fig:o3o2_fesc} that a large O32 index is still 
a {\it necessary} condition for a significant \fesc, but that not all LAEs with large O32 values 
are Lyman continuum leakers. This implies that there may be a further additional physical 
property that must govern whether a LAE is a leaker. However, our examination of the full 
range of spectral diagnostics and the ionizing radiation field respectively shown in 
Figures \ref{fig:ews_ewlya}, \ref{fig:o3o2r23}, and \ref{fig:xi_Muv} reveals no fundamental distinction 
between LAE leakers and non-leakers. This follows a fundamental result we first introduced 
in \citetalias{fletcher2019} of this series, namely the puzzling dichotomy of LyC detections 
in the overall LACES sample.

As metal-poor, compact star-forming systems, LAEs are likely being seen in an early phase
of their evolution, providing abundant ionizing photons to explain their large O32 indices.
This would result in physical conditions that allow LAEs to leak LyC photons more frequently
than LBGs (\citetalias{fletcher2019}; see also \citealt{steidel2018}).
A possible explanation for the dichotomy presented in \citetalias{fletcher2019} further 
defined via the absence of any line diagnostic to separate leakers and non-leakers 
in the present analysis, is anisotropic leakage. In this hypothesis, the LACES sample would 
represent a fairly homogeneous sample, in terms of its spectroscopic properties and hardness 
of the radiation field, but the primary distinction between LyC-LAEs and noLyC-LAEs would 
be viewing angle. This could be considered as a less extreme version of the original 
density-bound nebula case discussed by \citet{NO2014} whereby the system is only 
partially porous to LyC radiation.

However, one important objection to a geometrical explanation is the fact that
LyC leakers and non-leakers also have similar EW(\Lya) distributions 
\citepalias{fletcher2019}.
If the paths of \Lya\ and LyC photons are similar, one might expect \Lya\ fluxes 
to be similarly diminished for the non-leakers.  
Indeed, leakages of both \Lya\ and LyC photons are indicated to be modulated
by the covering fraction of the optically thick neutral gas
(e.g. \citealt{reddy2016_fcov,chisholm2018,steidel2018}).
This point was discussed with the available \Lya\ data in Sections 5.1 and 6.1
of \citetalias{fletcher2019} 
where correlations between \fesc\ and EW(\Lya) and its velocity were presented. 
By construction, all LACES targets must have prominent \Lya\ emission so sources 
with obscured \Lya\ will be absent, possibly weakening any expected trends.

While \Lya\ photons could preferentially escape along the same holes in the neutral
medium as LyC photons, due to the resonant nature of the line, \Lya\ photons can 
also escape after experiencing several scatterings. This would also explain why some 
LyC leakers present a higher escape fraction of \Lya\ photons than that of LyC photons
(e.g. \citealt{verhamme2017}). 
Moreover, the previous studies at high-$z$ correlating the \Lya\ and LyC leakages 
with the covering fraction mostly investigate stacked LBGs with a weak \Lya\ emission
(up to EW(\Lya) $\sim 45$\,\AA), and hence in a low \fesc\ range 
(e.g. \citealt{reddy2016_fcov,steidel2018}).
The correlation between \Lya\ and LyC found by stacking does not require that 
every \Lya\ emitter is necessarily a LyC emitter. 
For example, \citet{japelj2017} investigate the LyC visibility of $z=3-4$ 
star-forming galaxies, most of which are \Lya\ emitters, finding that none of them
present significant emission of LyC radiation.
Conversely, \Lya\ could be weakened from LyC emitters due to 
a spatial variation of neutral hydrogen column density across an object
(e.g. \textit{Ion1}; \citealt{vanzella2012,ji2019}; see also \citealt{erb2019}
for suppression mechanisms of \Lya\ emission).
Thus while LyC leakage is broadly correlated with \Lya\ properties, it is unclear 
whether the tight correlation holds between the observed escape fractions of 
\Lya\ and LyC emission on an individual basis.
Further data, e.g. higher resolution spectra sampling
the \Lya\ emission line profile for our sources, would be advantageous 
to investigate these possibilities.
If a narrow peak of \Lya\ emission is identified at systemic velocity, 
as seen in the other known LyC emitters of \textit{Sunburst}
\citep{rivera-thorsen2017,rivera-thorsen2019}
and \textit{Ion2} and \textit{Ion3} \citep{vanzella2019_ion2}, 
the detection of LyC emission would be attributed to 
a clear ionized channel along our line of sight.

Figure \ref{fig:o3o2_xi_ewo3} shows the relationship between O32 and \xiion\ versus
the EW of \OIII\ for the LACES sample, lower redshift extreme emission line galaxies
(EELGs, \citealt{tang2019}) and continuum-selected galaxies from the MOSDEF 
survey \citep{sanders2019}. We can see that continuum-selected galaxies and 
less massive EELGs are  similarly distributed in both panels.
Since the EW(\OIII) is an approximate measure of the age of the most recent star
formation activity (i.e. the specific SFR) as well as the ISM ionization and metallicity,  
the overall trends indicate younger stellar populations in a more highly ionized,
lower metallicity environment have both a larger O32 and harder \xiion\ 
as shown by \citet{tang2019}.

However, despite these strong correlations, the LACES 
LAEs, both leakers and non-leakers, fall above the sequence, 
presenting an enhanced O32 for a given EW(\OIII).
Although different ISM conditions may partially explain
the apparent difference in O32 between galaxies selected by \Lya\ and optical
emission lines, such an enhancement in O32 would support some version of 
the density-bound or porous nebula hypothesis \citep{NO2014}.
If the enhanced O32 is true and confirmed for both the leakers and non-leakers, 
the primary distinction between the two population would be 
an independent physical property, such as viewing angle. 
Indeed, a local strong LyC leaking source, J1154+2443, with \fesc\ $=0.46$ 
present almost the same large values of O32, \xiion\ and EW(\OIII) as seen in the 
composite of noLyC-LAEs from our LACES sample \citep{schaerer2018}, 
implying that noLyC-LAEs could have a condition to emit LyC radiation, but the 
pathway is not along our line of sight.

Admittedly, it is hard to verify the viewing angle explanation 
directly with the current dataset. Conceivably examining \Lya\ profiles with higher 
spectral resolution than is currently available (e.g. \citealt{verhamme2015}) and/or 
the depth of interstellar absorption lines in the rest-frame UV wavelength 
(e.g. \citealt{heckman2011,reddy2016_fcov,chisholm2018}) 
might provide further evidence of the geometrical hypothesis.
Interestingly, deep composite UV spectra of LAEs are reported to present 
a tantalizing trend that LAEs {\it on average} show shallow interstellar absorption lines, 
i.e. low covering fractions of low-ionization gas, significantly lower than those seen in LBGs
\citep{jones2013_fesc,trainor2015,steidel2018}, 
although it is not known which of these individual LAEs present a direct LyC leakage.
Such an investigation for leakers and non-leakers over wider dynamic ranges of 
\fesc\ and EW(\Lya) would be useful to describe the origin
of the \fesc-dichotomy and hence how ionizing photons escape from galaxies.

Finally, in \citetalias{fletcher2019} we considered a spatial variation of the IGM 
transmission as a contributing factor to the leaker/non-leaker dichotomy noting 
the SSA22 field contains a proto-cluster at $z=3.1$. Conceivably the \HI\ gas distribution 
may be complex \citep{mawatari2017,hayashino2019}. However, no spatial differences 
were seen between the distribution of LyC leakers and non-leakers in \citetalias{fletcher2019}. 
We stress, however, that our current sample is too small for any significant clustering patterns
to be discerned and so we still consider this explanation of the dichotomy discussed in this paper a
plausible alternative. We plan to address this question via a LyC search from LAEs at lower
redshifts where the IGM opacity  and its variation is less important \citep{inoue2014}.

In summary, we have extended our analysis of the spectroscopic properties of the LACES 
sample of $z\simeq$3.1 LAEs from that presented in \citetalias{fletcher2019}. Specifically 
we have added measures of the O32 index (based on new Keck spectra sampling \OII\ emission) 
as well as of \xiion, the hardness of the radiation field. Although a strong O32 index is 
a necessary condition for escaping radiation, we find that both LyC leakers and non-leakers 
have similar O32 and \xiion\ values, suggesting that an additional physical property must 
govern whether escaping radiation can be detected with HST. 
Our results  support the hypothesis that most LACES LAEs are likely emitting LyC radiation 
through a porous interstellar medium but suggest that only a fraction are being viewed favorably 
by the observer as LyC leakers.

\acknowledgments

The W. M. Keck observations were carried out within the framework of Subaru-Keck
time exchange program, where the travel expense was supported by the Subaru 
Telescope, which is operated by the National Astronomical Observatory of Japan.
Some of the data presented herein were obtained at the W. M. Keck Observatory, 
which is operated as a scientific partnership among the California Institute of Technology, 
the University of California and the National Aeronautics and Space Administration. 
The Observatory was made possible by the generous financial support of the W. M. 
Keck Foundation. The authors wish to recognize and acknowledge the very significant 
cultural role and reverence that the summit of Maunakea has always had within the 
indigenous Hawaiian community. We are most fortunate to have the opportunity to 
conduct observations from this mountain.
We thank the anonymous referee for helpful comments and discussions that improved 
our manuscript.
R.S.E. acknowledges funding from the European Research Council (ERC) under the 
European Union's Horizon 2020 research and innovation programme 
(grant agreement No 669253).
B.E.R. acknowledges support from NASA program HST-GO-14747, 
contract NNG16PJ25C, and grant 17-ATP17-0034.
The Cosmic DAWN Center is funded by the Danish National Research Foundation.

%

\vspace{5mm}
\facilities{Keck I (MOSFIRE)}



\begin{longrotatetable}
\begin{deluxetable*}{lccccccccccc}
\tablecaption{Physical Properties of the MOSFIRE-Identified Sources and their Composites
\label{tbl:physical_properties}}
\renewcommand{\arraystretch}{1.4}
\tabletypesize{\scriptsize}
\tablehead{
\colhead{Obj.} &
\colhead{$M_{\rm UV}$} &
\colhead{EW(Ly$\alpha$)} &
\colhead{$z_{{\rm sys}}$} &
\colhead{$\Delta v_{{\rm Ly}\alpha}$} &
\colhead{EW([O\,{\sc iii}])} &
\colhead{EW(H$\beta$)} &
\colhead{[O\,{\sc iii}]$/$H$\beta$} &
\colhead{R23} &
\colhead{O32} &
\colhead{$\log\,\xi_{{\rm ion}}$} &
\colhead{$f_{{\rm esc}}$}
\\
\colhead{} &
\colhead{} &
\colhead{(\AA)} &
\colhead{} &
\colhead{(km\,s$^{-1}$)} &
\colhead{(\AA)} &
\colhead{(\AA)} &
\colhead{} &
\colhead{} &
\colhead{} &
\colhead{(Hz\,erg$^{-1}$)} &
\colhead{}
}
\startdata
M38
& $-21.5\pm 0.0$
& \nodata
& $3.2911$
& \nodata
& $487.1\pm 4.8$
& $93.9\pm 2.4$
& $5.1\pm 0.1$
& $7.0\pm 0.2$
& $2.8\pm 0.1$
& $25.42\pm 0.02$
& \nodata
\\
2132
& $-21.9\pm 0.0$
& $7^{+2}_{-2}$
& $3.0586$
& \nodata
& $384.3\pm 7.8$
& $51.0\pm 4.7$
& $7.5\pm 0.7$
& $10.5\pm 1.0$
& $2.4\pm 0.2$
& $24.91\pm 0.04$
& \nodata
\\
104037$^{(\dag\,S)}$
& $-21.3\pm 0.0$
& $36^{+3}_{-2}$
& $3.0650$
& $224.2$
& $791.3\pm 17.4$
& $96.3\pm 4.4$
& $8.2\pm 0.3$
& $9.6\pm 0.4$
& $5.9\pm 0.2$
& $25.52\pm 0.02$
& $0.13\pm 0.02$
\\
93564$^{(\dag\,G)}$
& $-21.2\pm 0.0$
& $59^{+6}_{-6}$
& $3.6770$
& $545.5$
& $1040.5\pm 33.7$
& $120.7\pm 11.6$
& $8.6\pm 0.8$
& $9.5\pm 0.9$
& $10.0\pm 0.9$
& $25.72\pm 0.05$
& $0.31\pm 0.03$
\\
104511
& $-20.6\pm 0.1$
& $26^{+4}_{-4}$
& $3.0645$
& \nodata
& $1351.7\pm 44.7$
& $173.8\pm 14.2$
& $7.8\pm 0.6$
& $8.2\pm 0.6$
& $19.7\pm 3.7$
& $25.70\pm 0.04$
& \nodata
\\
108679
& $-19.8\pm 0.1$
& $54^{+11}_{-10}$
& $3.1066$
& $335.1$
& $1598.8\pm 87.2$
& $<96.1$
& $>16.6$
& \nodata
& \nodata
& $<25.60$
& \nodata
\\
96688
& $-21.9\pm 0.0$
& $-1^{+1}_{-1}$
& $3.1107$
& \nodata
& $247.0\pm 5.0$
& $21.4\pm 4.4$
& $11.4\pm 2.4$
& $15.5\pm 3.2$
& $2.8\pm 0.2$
& $24.54\pm 0.08$
& \nodata
\\
99330
& $-19.9\pm 0.1$
& $52^{+7}_{-6}$
& $3.1057$
& $341.7$
& $1696.7\pm 116.5$
& $169.5\pm 18.7$
& $10.0\pm 0.7$
& $11.0\pm 0.8$
& $10.2\pm 1.2$
& $25.65\pm 0.05$
& \nodata
\\
109140
& $-19.5\pm 0.2$
& $65^{+18}_{-15}$
& $3.1090$
& \nodata
& $2428.9\pm 262.3$
& $<268.7$
& $>9.0$
& \nodata
& \nodata
& $<25.75$
& \nodata
\\
86861$^{(\dag\,G)}$$^{\star}$
& $-21.4\pm 0.0$
& $81^{+3}_{-3}$
& $3.1054$
& $217.6$
& $295.4\pm 13.3$
& $30.3\pm 1.6$
& $9.7\pm 0.7$
& $9.7\pm 0.7$
& $>10.0$
& $25.18\pm 0.05$
& $0.46\pm 0.05$
\\
97030
& $-19.2\pm 0.2$
& $26^{+11}_{-9}$
& $3.0735$
& \nodata
& $853.7\pm 82.5$
& $91.0\pm 14.5$
& $9.4\pm 1.0$
& $10.8\pm 1.1$
& $6.6\pm 1.1$
& $25.72\pm 0.09$
& \nodata
\\
92017
& $-19.0\pm 0.3$
& $>90$
& $3.1070$
& $143.8$
& \nodata
& \nodata
& $>6.0$
& \nodata
& \nodata
& $<25.89$
& \nodata
\\
106500
& $-19.1\pm 0.2$
& $90^{+40}_{-30}$
& $3.0581$
& \nodata
& $2201.7\pm 437.2$
& $559.8\pm 150.5$
& $3.9\pm 0.5$
& $3.9\pm 0.5$
& $>9.5$
& $26.00\pm 0.10$
& \nodata
\\
104097
& $-20.8\pm 0.1$
& $-1^{+8}_{-6}$
& $3.0674$
& \nodata
& $173.1\pm 7.1$
& $48.9\pm 2.7$
& $3.5\pm 0.2$
& $5.3\pm 0.3$
& $2.0\pm 0.2$
& $25.00\pm 0.03$
& \nodata
\\
102334
& $-20.2\pm 0.1$
& $30^{+5}_{-4}$
& $3.0902$
& $203.0$
& $301.8\pm 21.0$
& $54.6\pm 12.5$
& $5.5\pm 1.3$
& \nodata
& \nodata
& $25.34\pm 0.09$
& \nodata
\\
94460$^{(\dag\,S)}$
& $-19.9\pm 0.1$
& $51^{+9}_{-8}$
& $3.0723$
& $157.5$
& $384.9\pm 21.6$
& $45.8\pm 11.6$
& $8.4\pm 2.1$
& $8.4\pm 2.1$
& $>17.3$
& $25.45\pm 0.11$
& $0.33\pm 0.02$
\\
102826
& $-21.1\pm 0.0$
& $-4^{+4}_{-3}$
& $3.0714$
& \nodata
& $94.4\pm 4.4$
& $17.0\pm 3.0$
& $5.5\pm 1.0$
& $9.0\pm 1.6$
& $1.6\pm 0.1$
& $24.62\pm 0.07$
& \nodata
\\
107585
& $-19.3\pm 0.2$
& $25^{+11}_{-8}$
& $3.0895$
& \nodata
& \nodata
& \nodata
& $3.1\pm 0.8$
& $3.1\pm 0.8$
& $>7.8$
& $25.85\pm 0.11$
& \nodata
\\
110896
& $-20.7\pm 0.1$
& $9^{+2}_{-2}$
& $3.0644$
& \nodata
& \nodata
& \nodata
& $>4.3$
& \nodata
& \nodata
& $<24.79$
& \nodata
\\
89114
& $-19.5\pm 0.2$
& $40^{+10}_{-9}$
& $3.0832$
& \nodata
& $1219.7\pm 104.9$
& $<135.5$
& $>9.0$
& $>9.0$
& $>8.6$
& $<25.33$
& \nodata
\\
99415
& $-19.2\pm 0.2$
& $62^{+21}_{-16}$
& $3.0972$
& \nodata
& $1019.7\pm 233.8$
& $193.9\pm 62.9$
& $5.3\pm 1.0$
& $5.3\pm 1.0$
& $>6.6$
& $25.59\pm 0.11$
& \nodata
\\
97081
& $>-18.7$
& $>208$
& $3.0762$
& $178.0$
& $>1890.0$
& $>197.2$
& $9.6\pm 2.7$
& $9.6\pm 2.7$
& $>10.4$
& $>25.52$
& \nodata
\\
90428
& $-19.4\pm 0.2$
& $51^{+10}_{-8}$
& $3.1037$
& $230.1$
& $>1743.6$
& \nodata
& $>5.2$
& \nodata
& \nodata
& $<25.46$
& \nodata
\\
93474
& $-19.3\pm 0.2$
& $50^{+16}_{-13}$
& $3.0702$
& $363.1$
& $2056.9\pm 383.3$
& $<428.3$
& $>4.8$
& $>4.8$
& $>7.0$
& $<25.55$
& \nodata
\\
85165
& $-21.5\pm 0.0$
& $35^{+2}_{-2}$
& $3.0996$
& \nodata
& $56.8\pm 8.3$
& $18.3\pm 4.8$
& $3.1\pm 0.9$
& $3.1\pm 0.9$
& $>2.3$
& $24.88\pm 0.10$
& \nodata
\\
92616$^{(\dag\,G)}$
& $-19.5\pm 0.2$
& $49^{+13}_{-11}$
& $3.0714$
& $253.3$
& \nodata
& \nodata
& $>6.2$
& \nodata
& \nodata
& $<25.68$
& $0.60\pm 0.09$
\\
104147
& $-19.4\pm 0.2$
& $24^{+8}_{-6}$
& $3.0994$
& $389.8$
& $371.6\pm 60.2$
& $<35.2$
& $>10.5$
& $>10.5$
& $>7.9$
& \nodata
& \nodata
\\
92219
& $-19.5\pm 0.2$
& $115^{+22}_{-18}$
& $3.1008$
& $182.8$
& \nodata
& \nodata
& $>7.4$
& $>7.4$
& $>5.7$
& \nodata
& \nodata
\\
93004
& $-19.4\pm 0.2$
& $38^{+13}_{-11}$
& $3.1127$
& $212.8$
& $1050.6\pm 170.9$
& $<221.3$
& $>4.7$
& \nodata
& \nodata
& $<25.49$
& \nodata
\\
92235
& $-20.2\pm 0.1$
& $29^{+6}_{-5}$
& $3.0713$
& \nodata
& $189.9\pm 17.7$
& $42.3\pm 12.2$
& $4.5\pm 1.3$
& $5.8\pm 1.7$
& $3.5\pm 0.6$
& $25.13\pm 0.11$
& \nodata
\\
97254
& $-18.9\pm 0.3$
& $68^{+24}_{-19}$
& $3.0712$
& $251.8$
& $745.1\pm 94.3$
& $109.0\pm 29.1$
& $6.8\pm 1.5$
& $6.8\pm 1.5$
& $>9.6$
& $25.49\pm 0.13$
& \nodata
\\
97176
& $-19.8\pm 0.1$
& $60^{+15}_{-12}$
& $3.0751$
& $219.2$
& $288.9\pm 27.7$
& $<40.7$
& $>7.1$
& $>8.5$
& $5.1\pm 1.2$
& $<25.10$
& \nodata
\\
103371
& $>-18.7$
& $151^{+72}_{-46}$
& $3.0894$
& $-20.5$
& $>1665.5$
& $>174.0$
& $9.6\pm 2.8$
& $9.6\pm 2.8$
& $>9.9$
& $>25.47$
& \nodata
\\
89723
& $-20.5\pm 0.1$
& $99^{+26}_{-20}$
& $3.1113$
& $259.6$
& \nodata
& \nodata
& $>9.1$
& $>9.1$
& $>8.8$
& \nodata
& \nodata
\\
110290
& $-19.3\pm 0.2$
& $56^{+20}_{-15}$
& $3.1088$
& $103.6$
& $653.8\pm 170.4$
& $198.8\pm 71.7$
& $3.3\pm 1.1$
& $3.3\pm 1.1$
& $>4.5$
& $25.53\pm 0.13$
& \nodata
\\
93981
& $>-18.7$
& $>71$
& $3.0766$
& \nodata
& $731.1\pm 94.6$
& $220.5\pm 51.8$
& $3.3\pm 0.7$
& \nodata
& \nodata
& $>25.74$
& \nodata
\\
105937$^{(\dag\,S)}$
& $-20.2\pm 0.1$
& $31^{+8}_{-7}$
& $3.0666$
& $155.5$
& $104.1\pm 15.4$
& $<65.9$
& $>1.6$
& $>1.6$
& $>3.4$
& $<25.62$
& $0.32\pm 0.07$
\\
91055
& $>-18.7$
& $>76$
& $3.0818$
& $254.9$
& $1041.0\pm 173.1$
& $218.9\pm 80.1$
& $4.8\pm 1.5$
& $4.8\pm 1.5$
& $>4.6$
& $>25.53$
& \nodata
\\
107677
& $>-18.7$
& $>109$
& $3.0679$
& $47.2$
& \nodata
& \nodata
& $>1.2$
& \nodata
& \nodata
& \nodata
& \nodata
\\
95217
& $-19.0\pm 0.3$
& $81^{+49}_{-32}$
& $3.0668$
& \nodata
& $1020.4\pm 372.0$
& $<526.3$
& $>1.9$
& $>1.9$
& $>2.8$
& $<25.73$
& \nodata
\\
97128
& $>-18.7$
& $>81$
& $3.0725$
& \nodata
& $>751.8$
& $>552.5$
& $1.4\pm 0.5$
& \nodata
& \nodata
& $>25.96$
& \nodata
\\
101846$^{(\dag\,S)}$
& $>-18.7$
& $>147$
& $3.0565$
& $232.1$
& $>166.6$
& \nodata
& $>1.5$
& $>1.5$
& $>1.9$
& \nodata
& $0.42\pm 0.09$
\\
90675$^{(\dag\,G)}$
& $>-18.7$
& $>61$
& $3.1110$
& $-3.6$
& $<128.9$
& $252.5\pm 32.7$
& $<0.4$
& \nodata
& \nodata
& $>26.41$
& $0.39\pm 0.11$
\\
\hline
\multicolumn{12}{c}{Composite Spectra}
\\
\hline
LyC-LAEs
& $-20.1\pm 0.6$
& $44\pm 11$
& \nodata
& \nodata
& $600.0^{+293.5}_{-206.8}$
& $99.2^{+32.2}_{-31.1}$
& $7.0^{+2.4}_{-2.4}$
& $7.6^{+2.6}_{-2.6}$
& $10.5^{+3.0}_{-3.0}$
& $25.65^{+0.11}_{-0.18}$
& $0.35\pm 0.14$
\\
noLyC-LAEs
& $-19.4\pm 0.6$
& $64 \pm 27$
& \nodata
& \nodata
& $1067.0^{+182.0}_{-157.2}$
& $126.8^{+43.3}_{-34.9}$
& $7.7^{+1.7}_{-1.4}$
& $8.4^{+1.9}_{-1.6}$
& $12.7^{+7.5}_{-6.7}$
& $25.50^{+0.09}_{-0.09}$
& $<0.005$$^\ddag$
\\
LBGs
& $-21.3\pm 0.5$
& $0.5\pm 4.$
& \nodata
& \nodata
& $266.8^{+78.8}_{-31.9}$
& $40.3^{+17.7}_{-15.1}$
& $7.0^{+1.2}_{-0.9}$
& $10.5^{+1.7}_{-1.7}$
& $2.0^{+0.7}_{-0.5}$
& $25.05^{+0.05}_{-0.08}$
& $<0.005$$^\ddag$
\\
\enddata
\tablecomments{%
Upper/Lower-limits represent the $3\sigma$ values.
For the EW measurements of \OIII\ and \Hb, we use the HST/F160W photometry
in determining the continuum level (see \citetalias{fletcher2019}).
No constraint on EW is thus given if the object lacks the F160W coverage.
($\dag$) LyC leaking candidates from \citetalias{fletcher2019}.
The $G$ and $S$ denotes the Gold and Silver sample, respectively.
($\star$) The single LAE-AGN in the LACES sample.
($\ddag$) The upper-limit is drawn from the composite of all the non-detections 
including both LAEs and LBGs \citepalias{fletcher2019}.
}
\end{deluxetable*}
\end{longrotatetable}

\bibliographystyle{aasjournal}{}
\bibliography{Refs_paper.bib}{}



\end{document}